\documentclass[preprint,12pt]{aastex}
\begin{document}

\newcommand{\cms}{\,{\rm cm$^{-2}$}\,}
\newcommand{\kms}{\,{\rm km\,s$^{-1}$}\,}
\newcommand{\kmsmpc}{\,{\rm km\,s$^{-1}$\,Mpc$^{-1}$}\,}
\newcommand{\ev}{\,{\rm eV\ }}
\newcommand{\kel}{\,{\rm K\ }}
\newcommand{\etal}{{ et~al.~}}
\newcommand{\ergs}{\,{\rm erg\,s$^{-1}$}\,}
\newcommand{\ergscm}{\,{\rm erg\,s$^{-1}$\,cm$^{-2}$}\,}


\title{Chandra Observations of Galaxy Cluster Abell 2218}

\author{Marie E. Machacek\altaffilmark{1}, Mark W. Bautz, 
        Claude Canizares}
\affil{Center for Space Research, Massachusetts Institute of Technology,
       Cambridge, MA 02139}

\author{Gordon P. Garmire}
\affil{The Pennsylvania State University, University Park, PA USA}

\altaffiltext{1}{on leave from Department of Physics, Northeastern University,
   Boston, MA 02115}


\begin{abstract}
  We present results from two observations 
(combined exposure of $\sim 17$ ksec)
of galaxy cluster Abell 2218 using the Advanced CCD Imaging 
Spectrometer on board the Chandra X-ray Observatory that were taken on 
October 19, 1999.  Using a Raymond-Smith single temperature plasma 
model corrected for galactic absorption we find a mean cluster temperature 
 of $kT = 6.9 \pm 0.5$~kev, metallicity of $0.20 \pm 0.13$ 
(errors are $90\%$ CL) and rest-frame luminosity in the $2$-$10$~kev 
energy band of $6.2 \times 10^{44}$~\ergs in a $\Lambda$CDM cosmology with 
$H_0=65$~\kmsmpc.  The brightness distribution within
$4'.2$ of the cluster center is well fit by a simple spherical beta model 
with core radius $66''.4$ and $\beta = 0.705$.  High resolution Chandra data 
of the inner $2'$ of the cluster show the x-ray brightness centroid displaced
$\sim 22''$ from the dominant cD galaxy and the presence of azimuthally 
asymmetric temperature variations along the direction of the cluster mass 
elongation.  X-ray and weak lensing mass estimates are in good agreement for 
the outer parts ($r > 200h^{-1}$) of the cluster; however, in the core 
the observed temperature distribution can not reconcile the x-ray and strong 
lensing mass estimates in any model in which the  
intracluster gas is in thermal hydrostatic equilibrium.
Our x-ray data are consistent with a scenario in which recent merger
activity in A2218 has produced both  significant non-thermal pressure
in the core and substructure along the line of sight; each of these
phenomena probably contributes to the difference between lensing and x-ray
core mass estimates.
\end{abstract}

\keywords{galaxies: clusters: general -- galaxies: clusters: individual 
(A2218) -- galaxies: intergalactic medium -- x-rays: galaxies}


\section{Introduction}
\label{sec:introduction}

Galaxy clusters are the largest gravitationally bound systems known. 
Within our current hierarchical picture of structure formation they 
form quite late at the knots and intersections of overdense filaments 
through mergers of subcluster sized gas clouds and galaxies, and these 
mergers may well, in many cases, be ongoing. Investigations of the mass 
distributions and merger dynamics of these clusters provide insight into the 
amount and nature of the dark matter expected to dominate the gravitational 
potential, and may elucidate the complex gas dynamics and energy transfer 
mechanisms of the merger process itself. After the internal structure and 
dynamical state of the cluster are well understood they can be used  
to deduce other cosmological parameters. For example, a comparison of the 
x-ray surface brightness to the reduction in the 
cosmic microwave background (CMB) due to inverse Compton scattering of CMB 
photons off the hot intra-cluster gas can be used to infer the value of 
the Hubble parameter $H_0$ (Sunyaev \& Zeld'ovich 1972, 1980; 
Birkinshaw \& Hughes 1994).  Thus galaxy clusters serve as important 
laboratories to constrain and expand upon cosmological models of structure 
formation.

Fortunately the large mass and relatively low redshift of galaxy clusters
also gives us the means to unravel the mystery of their complex dynamics. 
Observations of the dispersion velocities of galaxies probe the cluster 
gravitational potential in which they move. Details of the galaxy velocity 
distribution may signal recent galaxy-galaxy interactions and merging 
activity.  In many of these clusters gravitational lensing, both strong and 
weak, provides a direct measure of the total projected mass, gas plus dark 
matter, within the cluster.  Strong lensing measures the total projected mass
at small radii, i.e within the critical radius of the lensed arc, while weak 
lensing is more sensitive to the mass distribution out to larger radii.  
Thermal x-ray emission from the hot intra-cluster gas measures directly the 
gas density and temperature distributions within the cluster. This hot gas 
is confined by and expected to trace the underlying three-dimensional 
gravitational potential. Under the assumption of thermal hydrostatic 
equilibrium these gas properties may be used to infer the cluster total mass
distribution. A comparison of x-ray to lensing projected total mass poses 
an important check on the validity of these hydrodynamical assumptions.  For
most dynamically-relaxed clusters, where the cluster contains a single 
dominant galaxy at its center, agreement is quite good. The x-ray surface 
brightness contours tend to be smooth and roughly spherical. The mass 
centroid determined by lensing and x-ray analyses coincide with each other, 
with the peak of the x-ray emission and also the position of the dominant 
galaxy (See for example, Allen 1998).  However, for clusters that appear 
more active dynamically the lensing mass may be as much as a factor of 
$2$ -- $3$ above the x-ray mass estimates (Miralde-Escude \& Babul 1995; 
Allen 1998). Possible explanations for such a significant disparity between 
the x-ray and lensing masses seem linked to the presence of recent merging 
activity within the cluster and include the possibility of complicated 
cluster temperature profiles (Makino  1996; Cannon \etal 1998), the 
presence of significant nonthermal pressure support in the cluster gas 
(Loeb \& Mao 1994, Miralda-Escude \& Babul 1995), or clumpiness 
along the line of sight (Bartelmann \& Steinmetz 1996).

Abell 2218 is a prototype galaxy cluster in this latter 
category. It is an Abell richness class $4$ cluster at redshift $z=0.175$ 
with an intricate array of gravitationally lensed arcs and more than $100$ 
arclets (Pello \etal 1992).  Its measured velocity dispersion, 
$1370^{+160}_{-120}$~\kms, is high for its x-ray temperature
(LeBorgne \etal 1992). Previous x-ray 
observations find mean cluster temperatures between $6.7$~keV 
(McHardy 1990; Cannon \etal 1998) and $7.2$~keV 
(Mushotzky \& Lowenstein 1997; Allen 1998), metallicities close to $0.2$, and 
a $2$--$10$ keV x-ray luminosity in a flat, standard cold dark matter 
cosmology ($H_0 = 50$~\kmsmpc, $\Omega_{m} = 1$ ) of 
$\sim 10^{45}$~\ergs (David \etal 1993; Allen 1998). 
There is evidence from ASCA data 
of a temperature decline for radii greater than a few arc minutes 
(Lowenstein 1997). Observations in both the optical and the x-ray suggest
recent dynamical activity. Natarayan \& Kneib (1996) found velocity 
dispersion profiles for galaxies in the cluster core that were inconsistent 
with  equilibrium.  In a recent photometric study of $49$ galaxies
in Abell 2218, Ziegler \etal (2001) find that older galaxies, as 
determined by their stellar populations, are not 
concentrated near the core, but are well mixed throughout the cluster. Such 
mixing is expected within a few dynamical times after a major 
merger event. They also find one galaxy (\#$1662$) in the core 
whose stars have an 
unusually high velocity dispersion suggesting that it may have been tidally 
stripped. Gravitational lensing analyses of the arcs and arclets in Abell 
2218 require multiply-peaked total mass concentrations within the cluster
(Kneib \etal 1996), while a nonparametric mass reconstruction 
using both strong and weak lensing data shows that the mass concentrations 
are offset from the light, i.e. the dominant cD and second brightest 
elliptical galaxies, respectively (AbdelSalam, Saha \& Williams, 1998). 
While early x-ray observations appeared to place the peak x-ray emission at
the location of the dominant cD galaxy (Birkinshaw \& Hughes 1994; 
Squires \etal 1996), more recent analyses using 
higher resolution ROSAT data confirm a more complex picture of the cluster 
core. Markevitch (1997) found evidence in gaussian smoothed ROSAT HRI data of 
multiple x-ray peaks significantly displaced from the cD galaxy, 
while Allen (1998) found, using an iterative procedure, that the centroid of 
the x-ray emission was also displaced from the cD galaxy toward the 
direction of the second brightest elliptical. Improvements in the
x-ray data and lensing analyses have failed, however, to reconcile the 
discrepancy of at least a factor $2$ between the projected mass 
distributions inferred by these two techniques. 
        
In this paper we present results of two  
observations ($\sim 11$~ksec and $\sim 6$~ksec, respectively) of Abell 2218 
by the Chandra X-ray Observatory. In Section \ref{sec:obs} we detail the 
observations and the data reduction and processing procedures.
In Section \ref{sec:results} we present our main results. We first discuss our
measurements of the global properties of the cluster. We then exploit 
the unprecedented spatial resolution of Chandra to focus on the structure of 
the cluster core. Finally, we determine the projected total 
mass distribution implied by the x-ray data and compare it to the lensing
results.  In Section \ref{sec:discuss} we interpret these findings in terms
of past merging activity within the cluster core and we summarize our 
results in Section \ref{sec:conclude}. All errors are $90\%$ confidence 
limits and coordinates are J2000.  Unless noted otherwise we state our
results in the currently preferred flat, low matter density $\Lambda$CDM 
cosmology ($\Omega_m=0.3$, $\Omega_{\Lambda}=0.7$, $h=0.65$) with 
Hubble parameter
$H_0=100h$~\kmsmpc. Thus for the cluster redshift $z=0.175$, $1''$ corresponds
to a distance scale of $2.08h^{-1}$~kpc. 
  

\section{Observations and Analysis}
\label{sec:obs}

Abell 2218 was observed twice for $11.4$ and $5.9$~ksec, respectively, by
the Chandra X-ray Observatory (CXO) on October 19, 1999. The data were 
obtained using the Advanced CCD Imaging Spectrometer (ACIS, 
Garmire \etal 1992; Bautz \etal 1998) with a focal plane 
temperature of $-110^\circ$~C throughout the observations.  
All four of the front-illuminated 
CCDs (I0-3) in the $2 \times 2$ imaging array (ACIS-I) were active in both 
exposures. In addition CCD(s) S2 (S2, S3) from the $1 \times 6$ 
spectrometer array  were active for the longer (shorter) exposure, 
respectively.  However, the telescope pointing placed the bulk of the 
cluster emission on CCD I-3. Thus in this analysis we use only data from the 
ACIS-I array for imaging and CCD I3 for spectral analysis. Each CCD is a 
$1024 \times 1024$ pixel array where each pixel subtends 
$0''.492 \times 0''.492$ on the sky. 

The data were ground reprocessed on 27 September 2000 by the Chandra X-ray 
Center (CXC) resulting in better quiescent background rejection, an improved 
aspect solution, and better modeling of instrument gain and quantum 
efficiency, including its position dependence within the chip. This latter 
modeling of quantum efficiency 
(non)uniformity partially corrects for the charge transfer inefficiency 
induced along the chip read-out direction for front-illuminated CCDs believed
to have been caused by exposure to low energy protons during the passage of 
the telescope through the Earth's radiation belts shortly after launch 
(Prigozhin \etal 2000).  For our data this 
is particularly important not only because the ACIS-I array is 
front-illuminated, but also because the telescope orientation placed the 
axis of the mass elongation of the cluster, along which we might 
expect important physical effects to appear, close to the read-out direction 
of the chip.  Thus we used an earlier release of the data only to 
check for the qualitative consistency of our results and the efficacy of 
cleaning and analysis algorithms. The results presented here are based 
solely on the reprocessed data sets using the most recent instrumental 
corrections. We used throughout the CXC Chandra Interactive Analysis of 
Observations analysis suite CIAO-$2.0$ for the imaging analyses and the High 
Energy Astrophysics software packages FTOOLS and XSpec to fit spectra.

Bad pixels were removed from each observation using the CXC provided 
observation dependent bad pixel files. We checked the light curves for each 
observation for periods of anomalously high background (flares) and found 
none. Cosmic ray contamination was minimized in the CXC reprocessed data
by rejecting ASCA event grades $1$, $5$ and $7$ and applying the algorithm 
{\tt acis-detect-afterglow} to remove residual flaring pixel events.  
Cosmic ray induced flaring pixels mimic point sources in the data. 
However, in high count rate data, the cleaning algorithm 
{\tt acis-detect-afterglow} has a tendency to 
overclean the data removing some signal. To check that our 
point source identifications are robust, we compared the results from 
analyses where the flaring pixels were identified by directly scanning the 
event list for each source candidate found in data not cleaned with 
{\tt acis-detect-afterglow} to results from the same observation 
after the cleaning algorithm was applied. We found that for our data sets 
{\tt acis-detect-afterglow} worked well, causing no significant loss of 
signal while removing all but $3$ events of the flaring pixel contamination. 
 Thus no further cleaning of the reprocessed data sets was required. 
We used CIAO-$2.0$ tools to combine the data sets for imaging and point 
source identifications.  For spectral analyses we analyzed the data sets 
jointly rather than combining the sets since the instrument files needed are 
observation dependent.  

In Figure \ref{fig_fullwpoints} we present a broadband 
($0.3$-$7$~keV), adaptively smoothed, exposure corrected 
Chandra/ACIS-I image of Abell 2218. The data have been binned by four such 
that an image pixel subtends $1.96$ arcsec$^2$ on the sky and adaptively
smoothed using CIAO tool {\tt csmooth}. Telescope vignetting 
and spatial efficiency variations  have
been corrected by means of an exposure map for the combined, two-observation 
data set. The exposure map was  generated using standard  CIAO tools, 
assuming the best-fit model spectral energy distribution for the integrated
cluster emission (see section \ref{sec:mean}, below).  We see that with 
Chandra's high spatial resolution  we resolve a large number of x-ray point 
sources within the field, i.e. $\sim 8'.4$ from the location of the cluster 
center, where cluster emission is also expected to be significant. 

In order to check the Chandra astrometry we ran the CIAO wavelet point
source detection algorithm {\tt wavdetect} (Dobrzycki \etal 2000; 
Freeman \etal 2001) on the broadband ($0.3$-$7$~keV) image of the cluster
field. With a {\tt wavdetect} significance parameter set to 
$10^{-7}$, the expected number of false detections for this field
is $0.4$, while we identified $52$ point source candidates.
A detailed discussion of these sources will be presented elsewhere. 
We compared our source 
list with the optical USNO-A$2.0$ catalogue (Monet 1998). Using a 
correlation radius of $0.''5$, we found four matches within a $14'$ radius
of the ACIS-I center.   
The cumulative 
probability of a single random association between these data is $0.014$; the 
probability that all four matches occurred by chance is 
$1.6 \times 10^{-9}$. Thus the astrometry is good to within the nominal 
$0''.5$ without further adjustments. All x-ray point source candidates were 
removed from the data for the analyses of the galaxy cluster diffuse emission
presented here. 

Background rates significantly affect model fitting of the surface brightness
profiles and spectral analyses, and thus estimates of the cluster mass. 
ROSAT images of Abell 2218 show that  
cluster emission extends to a radius $ \gtrsim 12'$ from the central cD 
galaxy (Cannon \etal 1998), so the ACIS-I field of view is too small to find 
a region in our
observations without significant cluster contamination from which to 
measure the background rate. We use instead the blank field 
background sets acisi\_B\_i0123\_bg\_evt\_060600.fits for the imaging array 
and acisi\_B\_i3\_bg\_evt\_060600.fits for CCD I-3 compiled by 
Markevitch (2001) and drawn for our observations using the script  
{\tt make\_acisbg\_CIAO2} supplied with the blank field data. We apply the 
same
cleaning algorithm to the data as was applied to the background sets. This 
results in the additional removal of low count rate periods at the beginning 
and end of the observations reducing the useful exposure 
times to $11,408$ and $5,704$~s, respectively.  Identical spatial and energy
filters are applied to source and background data so that the background
normalization is set by the ratio of exposure times.  We check this 
normalization by comparing high energy photon ($9$-$12$~keV) counts in 
the data, where background is expected to dominate, to that predicted by
the normalized blank field background sets and find that they differ 
by $5.2\%$.
We adopt this discrepancy as our relative uncertainty in the background 
level.
    

\section{Results}
\label{sec:results}

In this section we present the main results of our analysis. 
In  Section \ref{sec:mean} we determine the mean properties of the cluster as 
a whole and find that we are, in general, in excellent agreement with 
previous work.  In Section \ref{sec:core} we explore the 
complex structure of the inner $\sim 2'$ of the cluster core. In Section 
\ref{sec:mass} we use our data to determine the total projected mass as a 
function of radial distance from the dominant cD galaxy and compare our 
results with gravitational lensing analyses of the same regions. 
    
\subsection{Mean Cluster Properties}
\label{sec:mean}

\hbox{a) Spectroscopy \hfil}

To investigate the mean temperature and metallicity of the cluster we extract
spectra for each observation from events, after point sources are removed, 
that lie both on CCD I-3 and within a circular aperture with radius 
$5'.1$ (corresponding to a radius of $0.64h^{-1}$~Mpc) centered on the peak 
of the broadband x-ray emission.  We group the data to require a minimum of 
20 counts per spectral bin. We then jointly fit the spectral data from both 
observations  with a single temperature Raymond-Smith 
plasma model (Raymond \& Smith 1977) corrected for galactic absorption using 
Wisconsin photo-electric cross-sections (Morrison \& McCammon 1983).
For energies below $0.7$~keV the uncertainties in the currently available
 ACIS-I response functions are relatively large; while for energies above 
$9$~keV the data become background 
dominated. Thus we restrict our fit to the $0.7$--$9$~keV energy 
band resulting in a total sample size of $12,217$ ($6,169$) events in the 
$11.4$ ($5.7$)~ksec exposure, respectively. Once the redshift of the source
is fixed, an absorbed Raymond-Smith model has, in principle, four free 
parameters: the absorbing column density, temperature, metallicity, and 
normalization. We find that fitting all parameters simultaneously results in 
an anomalously high absorbing column density, with depressed temperature 
and increased metal abundance,  over that determined by a fit 
with similar reduced-$\chi^2$ in which the absorption column density
is fixed at its Galactic value.  Mazzotta \etal (2001) determined that this 
fit degeneracy was due to remaining calibration uncertainties in the 
ACIS-I CCDs and that reliable temperatures could be obtained by fixing the 
absorption column from previous measurements. We also adopt this 
strategy freezing the absorbing column at its Galactic value of 
$3 \times 10^{20}$~\cms (Stark 1984).   

Our best fit results using the blank field background sets discussed in 
Section \ref{sec:obs} are summarized in Table \ref{tab:meanprop}. We find a 
mean temperature and metal abundance for Abell 2218 of
$kT=6.9 \pm 0.5$~keV and $Z = 0.20 \pm 0.13$ (with reduced-$\chi^2=0.993$). 
These results are in 
excellent agreement with previous measurements by McHardy (1990) using GINGA 
data, by Mushotzky \& Lowenstein (1997) using an integrated 
ASCA spectrum, by Cannon, \etal (1998) using combined ASCA GIS2 and GIS3 
data, and by Allen (1998) using both SIS and GIS ASCA data.  We find the 
flux in the 
$0.7$--$9$~keV energy to be $8.5 \times 10^{-12}$~\ergscm for this aperture
with an implied rest-frame luminosity in the $2$--$10$~keV energy band of 
$6.2 \times 10^{44}$~\ergs ($h=0.65$).  
As a final check on the sensitivity of our
results to the blank field background normalization, we increase the 
normalization of the background by $5.2\%$ and $10.7\%$ and find 
$kT=6.7 \pm 0.5$~keV, $Z=0.18 \pm 0.13$ (with 
reduced-$\chi^2=1.002$) and $kT=6.5 \pm 0.5$~keV, $Z=0.15 \pm 0.12$
( with reduced-$\chi^2=1.020$), respectively. 
As expected, increasing the background normalization lowers 
the fit temperatures and metal abundances, but the dependence is weak and
all results agree within their $90\%$ confidence limits.

\hbox{b) Imaging \hfil}

We model the surface 
brightness distribution with a simple, spherically symmetric core-index (beta)
model of the form:
\begin{equation}
S_x(r) = S_{0x} \bigl (1 +  {r^2 \over r_c^2} \bigr )^{-3\beta+1/2}
\label{eq:sbeta}
\end{equation}
where $S_{0x}$ is the x-ray surface brightness in photon counts/arcsec$^2$
at the distribution centroid ($x_0$,$y_0$), $r$ is the projected 
radial distance from the centroid and $r_c$ is the core radius,  each
measured in arc seconds, and $\beta$ is a dimensionless index. If the plasma
is assumed to be isothermal, then this 
distribution can be inverted to infer a gas density distribution of the form  
\begin{equation}
\rho_g(r)= \rho_{g0} \bigl (1 + {r^2 \over r_c^2} \bigr )^{-3\beta/2}
\label{eq:gasdens}
\end{equation}
where $\rho_{g0}$ is the gas density at the center of the distribution.
We select a broad energy band, $0.3$--$7$~keV, for the analysis and 
choose an $8'.4 \times 8'.4$ field from the combined event lists for both
observations (total exposure $\sim 17$~ksec with $21,808$ total counts) for 
the image that is centered roughly at the visual peak of the broadband 
emission. This region is a compromise between the need to be big, since the 
beta model is sensitive to emission at larger radii, and the need to avoid 
the imaging array edges to eliminate spurious numerical effects from the 
fitting algorithm. The background is modeled as a constant taken to be the 
mean photon counts/arcsec$^2$ for the above energy range determined from the 
blank field background sets for the ACIS-I array and normalized, as before, 
by the ratio of exposure times. We allow the core radius $r_c$, index 
$\beta$, normalization $S_{x0}$, and the centroid parameters 
$x_0$,$y_0$ to vary. Their best-fit values are found through minimization of 
the log-likelihood Cash statistic appropriate for Poisson distributed data 
using the CIAO tool SHERPA.

As summarized in Table \ref{tab:betacore}, we find a core radius 
$r_c = 66''.4 (\pm 0.45)$, index 
$\beta=0.705$ $(+0.004,-0.003)$, normalization 
$S_{x0} = 0.965 (\pm 0.012)$~counts/arcsec$^2$ and the x-ray emission 
centroid located at  $\alpha = 16^h35^m52.4^s$($\pm 0.89''$),
 $\delta=+66^{\circ}12'34''.4$($\pm 0.84''$).   
The intervals in parentheses represent $90\%$ confidence limits.  If we 
relax our assumption of spherical symmetry and 
fit the region with an elliptical beta model instead, the ellipticity is 
small, $0.179 (\pm 0.009)$, implying an axis ratio of $0.82$, close to 
spherical and in agreement with previous work (Siddiqui 1995), with no 
significant change in the center of the distribution or $\beta$. 
The major axis of the large field beta model fit is oriented at an angle 
$4^\circ.6 \pm 2^\circ.8$ north of west. We note that this differs 
considerably from the direction of the orientation axis of the inner 
x-ray contours and 
mass elongation in the cluster core ($\sim 40^\circ$ north of west as seen in 
Figures \ref{fig_fullwpoints},~\ref{fig_corecsmoothimg}, and 
\ref{fig_bbaxisgrid}). Thus there is considerable isophotal twisting in the 
inner $2'$ of the cluster. The core radius of the elliptical fit also 
increases to $73''.1 (-0.4, +0.6)$ due to the assumed elliptical symmetry.  
However, in general, the deviations from spherical symmetry in the large field
are small so that our use of the spherical beta model to infer the global 
mass distribution should be adequate.  
Our values for the core radius and $\beta$ index in the spherical beta model 
fit are consistent at the $90\%$ confidence level with previous fits by 
Birkinshaw \& Hughes (1994) using Einstein IPC and HRI data and those by 
Squires \etal (1996) using ROSAT PSPC data, but are somewhat higher than the 
ROSAT HRI results ($r_c = 58''$, $\beta=0.63$) found by Markevitch (1997).
However, the $30$~ksec HRI exposure used by Markevitch would not have sampled 
well the outskirts of the cluster at radii much greater than $2'$.

More striking is the best-fit location of the 
x-ray emission centroid.   As shown in Figure 
\ref{fig_corecsmoothimg}, the central dominant 
cD galaxy is located at $\alpha=16^h35^m49.2^s$, $\delta= +66^\circ12'45''$ 
(Markevitch 1997).  Thus we find that the x-ray emission centroid is offset 
by $22''\pm 1''.2$ ($19''$ east and $11''$ south ) from the cD galaxy. 
This agrees within $5''$ with the x-ray emission centroid found using 
iterative methods by Allen (1998) in ROSAT HRI data and is in the same 
direction as, although $\sim 10''$ east of, the dominant peak in the total 
mass distribution needed by AbdelSalam, Saha \& Williams (1998) in a 
non-parametric analysis of the weak and strong lensing data.  Our 
measurement also agrees in magnitude with the x-ray emission peak offset 
from the cD galaxy found by Markevitch (1997),
 although our fit places the emission centroid closer to 
($\lesssim 4''$ east of) the galaxy \#$1662$ with anomalously high velocity 
dispersion (Ziegler \etal 2001) than to the bright face-on galaxy at the 
southern edge of the giant red arc. We find no evidence for lensed emission
from the giant arc, which was suggested by Markevitch (1997) as one of 
several possible interpretations of the x-ray structure seen in the ROSAT 
HRI image.

\subsection{The Central $2$ ArcMinutes}
\label{sec:core}

Let us next focus on the central $\sim 2'$ of the cluster core.
In Figure \ref{fig_corecsmoothimg} we show adaptively smoothed, 
exposure corrected x-ray images of the innermost core region of the cluster
for two energy bands, $0.3$-$2$~keV (upper left) and $2$-$7$~keV 
(upper right). As in Figure \ref{fig_fullwpoints} the data are binned by 
four, adaptively smoothed with minimum and maximum signal to noise thresholds
set at $3\sigma$ and $5\sigma$, respectively, and then corrected for 
telecope vignetting and spatial efficiency variations by means of an 
exposure map. Beneath each x-ray image we show the x-ray contours for that
image superposed on the Hubble image of the same field. 
We denote the main peaks in the x-ray images by S for the soft band
and H1 for the hard band (with a second possible hard emission peak at H2) 
and superpose these locations on all of the panels. The peaks are broad 
with adaptively determined smoothing scales of 
$7''$ for the soft (S) peak and 
$10''$ for the hard (H1) peak.  However, the 
images clearly show that the soft and hard emission peaks are 
displaced from each other. We summarize these offsets in Table 
\ref{tab:offset}. The peak of the soft emission lies 
$\sim 17.5''$ ($\sim 36h^{-1}$~kpc) to the 
south and east of the cD galaxy in the direction of the mass elongation and 
the location of the major mass concentration found in the 
nonparametric treatment of strong and weak lensing by 
AbdelSalam, Saha \& Williams (1998); while the 
hard emission peak H1 lies $\sim 20''$ ($\sim 42h^{-1}$~kpc)
to the north and west of the soft emission peak and close to (possibly just 
north of) the cD galaxy.
The contours in Figure \ref{fig_corecsmoothimg} are smooth but distorted and 
hint of possibly multi-peaked structures.
The coincidence of the soft band peak with the galaxy \#1662 in the optical 
image suggests that we may be seeing some contamination by galaxy emission.
We will comment on this further in Section \ref{sec:discuss}. 
However, the general elongation of the contours along the line containing
the two major mass concentrations as well as the distortions perpendicular
to this axis are features caused by the intracluster gas.
 
In Table \ref{tab:betacore} we present results of spherical beta model plus 
constant background fits for hard ($2$-$7$~keV), soft ($0.3$-$2$~keV) and 
broadband ($0.3$-$7$~keV) emissions 
determined from images restricted to the central $4'.2 \times 4'.2$
region of the cluster. 
The constant background 
is chosen as before to be the mean background taken from the ACIS-I 
blank field background sets restricted to the appropriate energy band and
normalized by the ratio of exposure times. We again allow the 
location of the centroid to be a free parameter in the fits to see if the 
energy band dependence of the peak emission shown in the images is reflected 
as well in a similar shift in the beta model centroids of the x-ray emission 
in each band. It is not. We find that the location of the best-fit centroid 
is remarkably stable with formal errors $\lesssim 2''$ in 
each case and agrees within the $90\%$ confidence limits for each fit. 
Since the location of the x-ray centroid is most sensitive to emission at 
large radii, this is further evidence that the disturbances in the cluster
are confined to the inner regions of the core.  The 
complicated image structure is reflected in the different, and 
anomalously high, core radii and $\beta$ indices required by the fits. This
inconsistency between beta model fits to the core and fits to the global 
cluster was also found by Markevitch (1997) in the ROSAT data.

We can gain further insight into the departure of the x-ray surface 
brightness and thus gas density distributions from the broadband 
large field beta model fit by investigating the beta model 
residuals for this region. We smooth these residuals 
using a fixed $2$-dimensional gaussian kernel with $\sigma=5''$ and in Figure
\ref{fig_bbmodelresiduals} overlay the  
contours from that image on the Hubble Space Telescope optical mosaic of the 
cluster core. The residuals show a deficit of x-ray emitting gas (negative
contours) extending between the two major mass concentrations and directed
initially perpendicular to the direction of the mass elongation. This 
deficit then arcs outward in a crescent like shape around the region 
containing the dominant cD galaxy. There is a steep gradient from negative
to positive contours just southeast of the primary mass concentration 
framing a large region of excess emission to the north and west of the 
dominant galaxy. A less prominent excess is also found near the second 
brightest elliptical. 

The images of Figures \ref{fig_corecsmoothimg} and  
\ref{fig_bbmodelresiduals} suggest variations in the 
x-ray emission along the direction of the mass elongation of the cluster.
In order to investigate this variation more quantitatively, we choose a set 
of rectangular ($252'' \times 4''.92$) grids aligned perpendicular to a line 
that contains the mean (beta model) centroid of emission and lies along the 
direction of the mass elongation (See Figure \ref{fig_bbaxisgrid}). 
In the top two panels of Figure \ref{fig_btprofiles} we show the surface 
brightness profile for each of the energy bands (soft, hard and broad) 
determined by projecting the photon counts in each grid box onto the axis 
line. Errors are $\sqrt N$ where $N$ is the number of counts in the grid 
box. The origin is set at the mean x-ray emission center determined from 
averaging over the beta models listed in Table \ref{tab:betacore} 
($\alpha=16^h35^m52.3^s$,$\delta=+66^{\circ}12'35''.2$). 
Positive angles are to the north and west. For reference, the location of 
the cD galaxy and second brightest elliptical galaxy, projected onto the 
line, are at $r=+20''.4$ and $r = -47''.1$, respectively. We denote 
these positions  on Figure \ref{fig_btprofiles} by solid and 
dotted vertical lines. The data are not background
subtracted. The mean background levels expected per box are $17.7$, $7.9$ and
$9.9$~counts for the broad, soft, and hard energy bands, respectively.  We
also plot (solid curve) the prediction of the large field beta model plus 
constant background on the 
grid for the broad ($0.3$--$7$~keV) energy band and show its residuals 
in the third panel of Figure \ref{fig_btprofiles}. Although the
large aperture beta model plus constant background provides a reasonable 
description of the broadband x-ray surface brightness given the statistical
accuracy of our data, weak deviations suggesting a possible 
bimodal structure and dynamical activity within the core are present within 
$100''$ of the beta model center. 
The soft band, with the exception of a possible excess near the 
location of the second brightest elliptical, seems to mirror closely the 
broadband distribution. 
The hard band data are, however, skewed to positive 
$r$ with a broad, nearly flat peak that extends over and beyond the location 
of the cD galaxy.
We note that the 
possible excess of photons observed in the soft band at $r \sim -45''$
lies near the location of 
the second mass concentration found by AbdelSalam, Saha \& Williams (1998) 
in their nonparametric fit to the gravitational lensing data. 

In the bottom panel of Figure \ref{fig_btprofiles} we plot the temperature
profile across the core along the same axis. However, for such short 
exposures we do not have enough photon counts per grid box to extract a 
reliable spectrum. Thus we group the grids such that each grouping has 
$\sim 3000$ total counts in the combined image data before extracting the 
spectra separately. We display these grid groups 
and the axis line overlaid on the adaptively smoothed broadband image of 
the cluster core in Figure \ref{fig_bbaxisgrid} for reference. 
We again jointly fit
the spectra from the two observations with a single temperature Raymond-Smith
plasma model fixing the absorption column at its galactic value 
($3 \times 10^{20}$~\cms) and the metal abundance at $0.2$ determined from 
the mean spectrum for the cluster.  The temperature profile shows clear, 
azimuthally asymmetric variation along the axis of the mass elongation. 
However, the temperature variations are mild.  The 
temperature remains close to its mean cluster value ($kT \sim 7$~keV) south 
and east of the emission center (for $-140'' \lesssim  r \lesssim 0$). It 
rises to $kT = 8.70^{+2.1}_{-1.5}$~keV across the region containing the 
flat  hard band 
emission peak ($0 \lesssim r \lesssim 60''$) and then falls abruptly to 
$kT = 5.5^{+1.2}_{-0.9}$~keV, well below the cluster mean temperature, for 
$+60'' \lesssim r \lesssim +140''$. Uncertainties are all $90\%$ confidence
limits. 

It is natural to ask whether the data are consistent
with shocks in the gas expected when two merging substructures collide.
We can crudely estimate the compression factor $\rho_1/\rho_0$ of the gas 
across such a shock front using the x-ray brightness data: 
\begin{equation}
{\rho_1 \over \rho_0} \approx \biggl ( {S_{x1} \over S_{x0}} \biggr )^{1/2}
\label{eq:densratio}
\end{equation}
where $\rho_1(\rho_0)$ are the downstream (upstream) gas densities, 
respectively, and $S_{x1}(S_{x0})$ are the corresponding x-ray photon counts 
per bin. Under the assumptions of an ideal gas and adiabaticity, the 
compression ratio, temperature ratio and gas velocities across the shock are 
then simply related through the shock jump conditions 
(Mihalas \& Mihalas 1984). Let $x = \rho_0/\rho_1$ denote the inverse of 
the compression factor. Then the downstream to upstream temperature ratio is 
\begin{equation}
{T_1 \over T_O} = {(\gamma + 1)x - (\gamma - 1)x^2 \over 
   (\gamma + 1)x - (\gamma -1)}
\label{eq:tempratio}
\end{equation}
where $\gamma = 5/3$ is the adiabatic index. In the rest frame of the 
shock the downstream to upstream velocity ratio is $u_1/u_0 = x$; while 
the shock velocity difference is 
\begin{equation}
u_0 - u_1 = \Biggl ( {kT_0 \over \mu m_p}(1-x) \biggl ({T_1 \over xT_0}-1 
\biggr ) \Biggr )^{1/2}
\label{eq:veldif}
\end{equation}
where $k$ is Boltzmann's constant, $\mu = 0.6$ is the mean molecular mass of 
the gas (assumed fully ionized), and $m_p$ is the proton mass. 
Upstream (downstream) Mach numbers $m_0$($m_1$) are given by
\begin{equation}
m_0^2 = 1 + {(\gamma + 1) \over 2\gamma} \biggl ({T_1 \over xT_0}-1 \biggr ) 
\,\,\,{\rm and}\,\,\,
m_1^2 = 1 - {(\gamma +1) \over 2\gamma} \biggl (1-{T_0x \over T_1} \biggr ).
\label{eq:mach}
\end{equation}

The most striking feature in Figure \ref{fig_btprofiles} is 
the asymmetric temperature drop observed for $r \ge 60''$.   
If this were to be interpreted as the upstream edge of a shock located at 
$r \approx 60''$, the measured 
temperature ratio $T_1/T_0 = 1.59$ with Equation \ref{eq:tempratio}
would imply a compression factor 
$x \approx 1.8$. Such a shock would not be strong with 
upstream(downstream) gas velocities and Mach numbers of 1920(1050)~\kms and 
1.59(0.69), respectively.  This is in agreement with simulations that 
predict Mach numbers for shocks from merger events in clusters to be 
$\lesssim 3$ (Sarazin 2001 and references therein). 
However, from Equation \ref{eq:densratio} we should also expect to see 
an x-ray surface brightness discontinuity $S_{x1}/S_{x0} \approx 3.3$ on the 
high temperature side of the shock ($r \lesssim 60''$) that is clearly not in 
the data. 
Furthermore, although the fractional changes in entropy and pressure
are formally positive across the discontinuity ($0.6^{+0.9}_{-0.6}$ and 
$0.6^{+1.0}_{-0.6}$, respectively), the errors are so large that we can 
not rule out the possibility that these properties are actually continuous.   
Thus, unless the required enhancement in x-ray surface brightness has been 
washed out due to projection effects, this temperature change is unlikely to 
be associated with a shock.  There does appear to be a  possible 
($\sim 1 \sigma$ ) excess of x-ray 
photons over that predicted by the beta model fit in three consecutive bins
($60'' < r < 70''$) in the cold region. This could represent slightly denser,
unmixed gas from the primary cluster core.  An alternative explanation  
could be that we are seeing evidence of a 
cold cloud or cold gas inflowing along a filament prior to merger.

Another statistically weak feature that might be correlated with a shock
is the sharp x-ray brightness enhancement seen at $r \simeq 45''$. 
If this feature is a shock and we ignore possible projection effects, 
the brightness discontinuity and 
Equation \ref{eq:densratio} imply a compression factor of $1.12$, 
a temperature ratio of $1.08$, and upstream (downstream) Mach numbers 
$1.08 (0.93)$, respectively. 
The $\sim 8\%$ temperature rise expected for such a shock
can not be detected with the current data set, nor can we definitively 
distinguish between a weak shock and a possible contact discontinuity at 
this location.

Thus our data show no evidence for strong shocks in the cluster core.   
However, much better temperature data are needed before the presence 
(or absence) of weak shocks can be definitively determined.

\subsection{Projected X-ray Mass Distribution}
\label{sec:mass}

One of the most interesting unresolved issues surrounding Abell 2218 is the
discrepancy between the measured mass extracted from x-ray 
data (assuming hydrostatic equilibrium) and that determined from the 
gravitationally lensed arcs and arclets. 
Although our data suggest that Abell 2218 is not fully relaxed,
it is still useful to follow the usual x-ray mass analysis prescription for 
the cluster in order to quantify the significance of the cluster's 
departure from the standard assumptions. 
Since the large field, broadband, spherical beta model seems 
a reasonable description of the surface brightness distribution for the 
cluster on large scales, we use it and Equations \ref{eq:sbeta} and 
\ref{eq:gasdens} to infer the gas density distribution. 
However, since the temperature variation across the core appears azimuthally 
asymmetric, we do not fit the temperature distribution with a radial profile.
Instead we assume isothermality and bracket the mass estimate allowed by our 
data by  
using both the mean cluster value $kT=6.9$~keV, determined
from the large aperture spectrum, and $kT=10.8$~keV, the $90\%$ upper limit
to the maximum temperature found from our more detailed 
analysis of the core. We note that the region of high temperature corresponds
to the region containing the brightest gravitationally lensed arcs.
However, we do not see gas temperatures as high as those proposed by 
Cannon \etal (1998) to resolve the discrepancy between the lensing and 
x-ray mass estimates for this region.  
Interestingly, Figure \ref{fig_bbmodelresiduals} shows that there is 
excess luminosity, implying excess mass, in this region over that predicted 
by the mean beta model. However, there are too few photon counts in the 
beta model residuals (see Figure \ref{fig_btprofiles}) to 
constrain a two-component model in this region.

An added complication to the x-ray mass analysis is the 
fact that most lensing analyses measure the total projected mass within a 
critical radius $b$ centered on the dominant cD galaxy while one of the 
main results of this work is that the center of the cluster gas density 
distribution is offset from the cD galaxy by $\sim 22''$. We take this 
offset into account when integrating the total mass within concentric 
cylinders of radius $b$ centered on the cD galaxy. 
Assuming that the gas is isothermal and in hydrostatic equilibrium,
the total mass density $\rho_{tot}$
is determined from the gas density $\rho_g$ by 
\begin{equation}
\rho_{total} = {kT \over 4\pi G\mu m_p}\nabla^2 \ln \rho_g^{-1}
\label{eq:hydrodens}
\end{equation}   
where $k$ is Boltzmann's constant, $T$ is the gas temperature, $\mu=0.6$ 
is the mean molecular mass for the ionized gas, $m_p$ is the proton mass, 
and $G$ is the gravitational constant. 
The total 
projected mass $M(<b)$ within the critical radius $b$
is then found by integrating Equation \ref{eq:hydrodens} over the volume of 
the cylinder of radius $b$, centered on the cD galaxy. 
As noted by Allen (1998) in a 
similar analysis of ROSAT data, the displacement of the gas mass center from
the lensing mass center  results in a lower projected mass than 
would otherwise have been determined from the given beta model had the 
lensing and gas mass centers coincided.  

In Figure \ref{fig_massprofile}
we compare the projected total mass $M(< b)$ within a radius $b$ of the 
cD galaxy obtained from our data for $kT=6.9$~keV (solid line) and 
$kT=10.8$~keV (dashed line) with $M(< b)$ obtained from strong lensing
analyses (Loeb \& Mao 1994; Kneib \etal 1995; Allen 1998), weak lensing 
analyses (Squires \etal 1996; Smail \etal 1997), and the minimum 
and maximum mass models from a nonparametric reconstruction method 
(AbdelSalam \etal 1998) that uses both strong and weak lensing data.
Note that the scatter in the lensing results at small radii is due to 
the nonuniqueness of the mass model used in the various analyses. Furthermore
the weak lensing results at large radii should be considered lower bounds
on the enclosed mass. Squires \etal (1996) estimated that 
residual cluster mass in their selected reference annulus would increase 
their measurement of the weak lensing mass within $r \lesssim 400h^{-1}$ by 
$20-60\%$ depending upon the assumed cluster mass model. 
However, since this correction factor is so 
highly model dependent, we show only their uncorrected measurements 
in Figure \ref{fig_massprofile} and Table \ref{tab:massratio}. 
We present our mass results in the sCDM ($\Omega_0=1$,$\Omega_{\Lambda}=0$,
$h=0.5$) cosmology for direct comparison with previous work. 
The results for the $\Lambda$CDM model that we have heretofore assumed  can 
be obtained by multiplying the mass by a factor $0.83$, the ratio of the 
$\Lambda$CDM ($h=0.65$) to sCDM ($h=0.5$) distance scales.

For clusters such as Abell 2218 where the mass 
structure is multi-peaked and may not follow the light, the uncertainties in 
the lensing mass determination can be large. However, all of the lensing 
results for the total projected mass within the critical radii of 
the brightest arcs ($b < 85$~kpc) lie well above the total mass determined
from the x-ray data.  
In Table \ref{tab:massratio} we list the ratio of the lensing to 
x-ray projected masses within selected radii $b$ centered on the dominant cD 
galaxy. The agreement between the weak lensing mass and the x-ray mass 
determined in our simple isothermal ($kT=6.9$~keV) spherical model is 
excellent at larger radii ($b \gtrsim 200h^{-1}$~kpc). However, the x-ray and
lensing mass estimates become more discrepant as $b$ decreases even if we 
use in our model the $90\%$ upper limit on the highest 
temperature measured for the core region as 
characteristic of all the mass in the projection cylinder. This latter 
estimate may be interpreted as a conservative upper bound on the projected 
x-ray mass since it not only represents the $90\%$ upper bound on the 
maximum temperature measured in the core, but by using this temperature for
all of the gas in the projection cylinder it also overestimates the 
contribution of the cooler gas at large cluster radii. 

\section{Discussion}
\label{sec:discuss}

The results discussed above in Section \ref{sec:results} strongly support
the hypothesis that Abell 2218 is not fully relaxed, especially in its core. 
Our first indication of dynamical activity within the cluster
is the displacement of the x-ray centroid from that of the dominant cD galaxy.
Numerical simulations demonstrate that complicated multipeaked differences 
between the total mass, light and x-ray emission in clusters signal recent or 
on-going merger activity  (see, for example, Bryan 1996; Evrard \etal 1996, 
Roetigger \etal 1998). Roetigger \etal (1998) showed that ram pressure 
induced by a slightly off-axis collision between a massive subcluster and
the primary mass concentration ($2.5$ times more massive) is sufficient to 
displace the gas distribution from that of the total mass and light.  
Furthermore, the subcluster's dark matter halo can maintain its identity for
several additional core crossings. On each encounter it continues to disturb 
the cluster core such that the relaxation of the cluster back to hydrostatic
equilibrium is slow. In related numerical studies Gomez \etal (2000) found
similar patterns of gas evolution in head-on collisions between a primary 
cluster and infalling subcluster with a $16:1$ mass ratio between the 
components provided that the gas content of the subcluster was 
sufficiently high.  The ratio of the primary to secondary mass 
concentrations in Abell 2218, as modeled through lensing analyses 
(Kneib \etal 1996) is $\sim 12:1$, intermediate between these extremes.
Thus we expect the general features of such a collision to 
be qualitatively similar. 

Upon closer examination of the cluster core we can probe in more detail
the dynamics of the proposed merger. The isophotal twisting and 
general elongation of the 
contours in Figure \ref{fig_corecsmoothimg} along the line containing the 
two major mass concentrations as well as the distortions perpendicular
to this axis can both be understood within this scenario. 
Gas dragged outwards by the subcluster potential from the region 
surrounding the primary mass concentration causes the general elongation 
along the merger axis.  
In simulations of mergers with $16:1$ ($2.5:1$) mass ratios between the 
components, this elongation of the brightness 
(and thus density) distribution caused by the gas response to the 
elongated gravitational potential persists for at least 
$1$~Gyr ($3$~Gyr) after the initial core crossing, respectively 
(Gomez \etal 2000; Roettiger \etal 1998).
Bulk flows generated by the initial merger event
produce a global circulation of the gas. Gas is driven perpendicular to the 
merger axis in places distorting the x-ray contours along the direction of 
these flows (Roettiger 1998). The departures from a simple isothermal
beta model displayed in the pattern of residuals seen in 
Figure \ref{fig_bbmodelresiduals} can be interpreted as a consequence of 
these gas flows.  
Within our merging hypothesis the lower density 
regions (negative residuals with respect to the mean beta model) would 
again correspond to regions where the
gas has been swept outward along the direction of the bulk velocity flows.  
The steep gradient
represents a region of compression of the gas near and beyond the dominant
galaxy. This could be caused partly by the compression of the gas due to
the shock produced in the initial collision, but more importantly by 
adiabatic compression of the gas when the global bulk circulation of 
the gas meets itself as the merger evolves (Roettiger 1998). If so this
region where the gas is compressed should correspond to a region of 
higher temperature due primarily to adiabatic 
heating of the gas when the velocity flows meet. This is consistent with 
the temperature profile shown in Figure \ref{fig_btprofiles} and would 
also account for the displacement of the peak in the hard emission seen in 
Figure \ref{fig_corecsmoothimg}b.
Such velocity flows may already have been measured in the Centaurus cluster
using ASCA data (Dupke \& Bregman, 2001).
The excess emission near the second brightest elliptical in Figure 
\ref{fig_bbmodelresiduals} could be due to gas dragged out of the 
primary core by the gravitational potential of the subcluster.

The brightness distribution and mild asymmetric temperature variations in
Figure \ref{fig_btprofiles}, although broadly averaged, provide important 
clues about the evolutionary stage of the merger.  We see no evidence for 
the bow shock and associated large temperature and brightness 
discontinuities expected in 
the early stages of a merger just after the primary collision. Rather the 
data are qualitatively consistent with simulations of a slightly off-axis 
merger $\sim 1-2$~Gyr after the initial interaction  
(Roettiger 1998, 1999).  In such a picture 
we expect the bow shock to be mildly supersonic so that it will 
have migrated out of our $\sim 1h^{-1}$~Mpc field in $\lesssim 1$~Gyr, the 
sound crossing time for this field. 
Regions of the gas have been adiabatically compressed and heated by the bulk 
velocities established in the early stages of the 
merger. The subcluster dark matter core has survived several additional core 
crossings, and with each crossing ``stirs'' the 
intracluster gas breaking the large scale velocity flows into smaller scale
turbulent eddies.  However, mixing of the gas (and the approach to 
hydrostatic equilibrium) is slow. Simulations show that large cold regions
of gas may still be present several gigayears after the first collision, 
which could explain the sudden
drop in temperature north and west of the cD galaxy and the azimuthally
asymmetric pattern of hot and cold regions found in our map. Furthermore 
by this time the extremes in temperature have been mitigated such that when
azimuthally averaged the cluster would appear largely isothermal, also 
in agreement with our data (Markevitch \etal 2000).

Although agreement is good between the x-ray and lensing estimates of the 
total projected mass at large radii($\gtrsim 200h^{-1}$~kpc) indicating that 
when averaged over large
scales the cluster appears largely isothermal and in hydrostatic equilibrium,
 the lensing analyses at the smallest scales ($\sim 40h^{-1}$~kpc) are 
consistently a factor $2 - 3$ larger than the x-ray mass estimates. 
We note, following various authors (Evrard 1990, Loeb \& Mao 1994, 
Miralda-Escude \& Babul 1995, Bartlemann \& Steinmetz 1996 ),  
that two distinct pheonomena associated with merging exacerbate the 
measured differences between the strong lensing and x-ray estimates in
the cluster core. First, the presence of non-thermal pressures, due
to bulk plasma flows, turbulence, or the merger-induced 
amplification of magnetic fields causes the x-ray method, which 
presumes hydrostatic equilibrium, to underestimate the mass. Second,
significant substructure along the line of sight is more likely
in dynamically active clusters and, if present, will cause
the lensing technique to overestimate the mass. We consider each of these 
phenomena in turn.

Due to the limited resolution of current numerical simulations, 
(Roettiger 1998,1999) it is difficult to determine from them whether 
nonthermal pressures alone can be as large as required by our data. 
These simulations do show, 
however, that both nonthermal kinetic pressures and magnetic field 
amplification can be significant during the intermediate stages of the 
merger ($\sim 1-2$~Gyr after the primary encounter) and that both effects 
tend to increase as the simulations are able to resolve smaller scales. 
Roettiger \etal (1998) found that when averaging over spheres of 
$0.4h^{-1}$~Mpc the
nonthermal kinetic pressure still reached values
$40 \%$ of the thermal pressure at $\sim 2$~Gyr after the primary encounter, 
twice that obtained at the same epoch for a sphere whose radius was $4$ times
larger ($r=1.5h^{-1}$~Mpc).  They concluded that neglect
of the dynamical pressure component during the first $\sim 2$~Gyr after
a major merger will cause a significant underestimate of cluster mass and 
that the underestimate will be most severe when the analysis is restricted
to the central regions of the cluster.  Thus it is possible 
that if we average over regions another order of magnitude 
smaller than those in the simulation, such as the $\sim 40h^{-1}$~kpc 
critical radius of the brightest 
lensed arc, that the dynamical pressure due to bulk and 
turbulent flows could reach 
values comparable to the thermal pressure of the gas in that 
region. We should expect, as seen in our analysis, that agreement 
between the x-ray mass 
and lensing mass estimates improve as we move to larger radii.  

Roettiger \etal (1999) also show that it is at this intermediate stage of 
the merger evolution when the bulk velocity flows become broken up into 
smaller scale turbulent eddies and amplification of localized magnetic fields
is maximal.  They find with their limited resolution 
($\sim 12.5$~kpc) that the magnetic energy 
on small scales is amplified by a factor $\sim 26$ over that in a 
nonmerging cluster and speculate that with better resolution simulations, the 
amplification of magnetic energies due to turbulent eddies on yet smaller 
scales could be much greater.  
New, high resolution observations of several dynamically 
active clusters by Chandra (Vikhlinin \etal 2001b, Sarazin 2001) indicate that
scales on the order of a few kiloparsecs are likely to be dynamically 
important, scales roughly an order of magnitude smaller than the highest 
resolution simulations by Roettiger \etal (1998,1999). For example, in 
Abell 3667 cold fronts with sharp edges were found where 
dramatic changes in temperature and x-ray surface brightness occur over 
$\sim 2$~kpc scales. The suppression of 
thermal conduction implied by the Chandra data for Abell 3667 argues for 
significant amplification 
of the magnetic field to strengths $\sim 10\,\mu$G over kiloparsec
scales, translating into modest (but still significant) nonthermal magnetic 
pressures  between $10$ and $20\%$ of thermal at the edge of the cold front
(Vikhlinin \etal 2001a). Measures of Faraday rotation in a 
sample of fully relaxed clusters, i.e. those with no strong radio sources, 
cooling flows or signs of recent merging activity, by Clarke \etal (2001)
imply magnetic field strengths $\sim 4\,\mu$G in a tangled magnetic field 
model with constant coherence length of $15$~kpc and $H_0=50$~\kmsmpc. 
They argue that magnetic fields 
could reach values as high as $\sim 8\,\mu$G for kiloparsec coherence 
lengths near the cluster cores. 

The presence of significant magnetic fields in the inner region of 
Abell 2218 is consistent with VLA data that show a weak, centrally 
located radio halo in the cluster core (Moffet \& Birkinshaw 1989; 
Giovanni \etal 1999). Such radio halos are usually found in clusters which 
show evidence of recent merging activity (Buote 2001; Buote \& Tsai 1996). 
The relativistic electrons responsible for the low frequency ($1.4$~GHz) 
radio emission observed in these halos are thought to be initially 
accelerated by the shocks produced
in the initial stages of a major merger. These electrons can then be 
reaccelerated by the subsequent turbulent eddies produced in the gas 
during the intermediate stage of the merger evolution after the shock has 
traveled out of the core.  The weakness of the radio halo found in 
Abell 2218 may be a consequence of the fact that we are observing the cluster 
at such an intermediate stage, $\gtrsim 1$~Gyr after the primary collision.
However, if we were to assume equipartition between 
the thermal, kinetic and magnetic energies in the innermost core of Abell 
2218, the data would require an average magnetic field strength within the 
critical radius of the brightest arc of $\gtrsim 50\,\mu$~G to reconcile the 
lensing and x-ray mass estimates.  
This would require amplification of the average magnetic field energy in
this region due to turbulent small scale eddies
by factors $\gtrsim 40$ over those expected near the cores of fully 
relaxed clusters and a factor $\gtrsim 2$ larger than found in 
existing simulations (Roettiger \etal 1999). 
 Thus, while it is likely that kinetic and magnetic pressures
are important during this stage of the merger evolution, we need both better
data and higher resolution simulations to determine whether quantitative 
agreement is possible solely due to the presence of these nonthermal 
pressures.  

Another possible contributor to the difference in lensing and x-ray masses
in Abell 2218 is the chance alignment of substructure along the line of 
sight.  Bartlemann and Steinmetz (1996) argued from simulations that 
ongoing merging activity, as
evidenced by substructure in clusters, significantly enhances
a cluster's ability to form multiple arc systems. This is consistent with
our conclusion that Abell 2218 is dynamically active. 
They further found that in the few simulated clusters in which the 
discrepancy between the projected lensing and x-ray masses was large 
(up to factors $\sim 2$, as is the case for our data) that the radial 
velocity distributions were bimodal indicating the infall of substructure 
along the line of sight. Studies of both the optical
(Zabludoff \& Mulchaey 1998) and x-ray (Mulchaey \& Zabludoff 1998) 
properties of a sample of nearby galaxy groups found that groups
with detectable x-ray emission behaved very much like low mass galaxy 
clusters. 
The x-ray emitting groups in this sample 
are bound systems with characteristic temperatures $\sim 1$~keV, a median 
harmonic radius $r_h = 0.82$~Mpc and virial masses 
ranging from $0.8 - 2 \times 10^{14}\,M_{\odot}$ (assuming $\Omega_0=1$, 
$h=0.5$). Zabludoff \& Mulchaey argue that the harmonic radius is a good 
approximation to $r_{500}$ where dynamical equilibrium is achieved for 
systems at this temperature. The virial mass is large compared to both the 
mass contained in x-ray gas ($\sim 6 \times 10^{12}\,M_{\odot}$) and in 
galaxies ($\sim 2 \times 10^{13}\,M_{\odot}$) and is distributed in a large, 
common halo with a bright elliptical galaxy located at its center. The x-ray
emission consists of two components, a compact (core radius $\sim 40$~kpc)
soft ($0.7 - 0.9$~keV) component thought to be associated with the central
elliptical galaxy and a fainter, extended component (core radius 
$\sim 200$~kpc) that is most likely emission from the intragroup gas.
These groups thus serve as useful models for the subclusters expected to 
participate in merging in a rich cluster environment.

If we assume a singular isothermal density profile  
($\rho \propto  r^{-2}$) for such a group infalling along the line of sight, 
the maximum contribution of the group to the 
projected total mass within the critical radius $b$ can be no larger than 
$M(<b) < (\pi/2)(b/r_v)M_v$ where $M_v$ and $r_v$ are the virial mass
and radius, respectively. 
For the brightest lensed arc of Abell 2218, $b=80$~kpc such that 
$b/r_v \sim 0.1$. Thus an infalling
group with mass in the range characteristic of the 
Zabludoff \& Mulchaey sample 
could account for as much as $30\% - 70\%$  of the discrepancy between the 
lensing and x-ray projected mass estimates within the brightest arc. Changing 
the density profile of the group to an NFW or Moore profile  
reduces the group contribution since each of these profiles 
is less concentrated ($\rho \propto r^{-1}$ or $\rho \propto r^{-1.5}$, 
respectively) in the core than the singular isothermal profile assumed above. 
X-ray emission from such an intervening group would most likely be 
observable in our Chandra data. The x-ray signature for such a group would be
enhanced soft emission, with 
a scale of $\sim 40$~kpc ($\sim 10''$), coincident with a bright elliptical 
galaxy, and with no corresponding feature in the hard band. It is interesting
to note that the peak of the soft emission in our data is nearly 
coincident with the elliptical galaxy \#1662 from Ziegler \etal (2001).
Thus an intervening group could possibly account 
for the displacement of the hard and soft band emissions seen in our data. 
 However, if galaxy \#1662 is interpreted as the
center of this group, as the x-ray emissions might imply, the $\sim 18''$  
offset of the group center from the dominant cD galaxy would also reduce the 
above estimate of the group's contribution to the projected mass.

Thus we conclude that recent merging within dynamically active clusters 
produces significant nonthermal pressures so that the standard (thermal
hydrostatic equilibrium) x-ray mass analyses tend to underestimate the 
cluster mass. In addition, the presence of substructure in these clusters  
increases the likelihood that the projection of an infalling group along the
line of sight will cause the strong lensing mass to be an overestimate thus 
enhancing the measured discrepancy. 
We find that the contribution from an intervening galaxy group along the 
line of sight may be significant in Abell 2218. However, it probably can not 
by itself reconcile the discrepancy between the lensing and x-ray projected 
mass estimates within the $40h^{-1}$~kpc radius of the brightest arc. 
Our data suggest that a full explanation of the lensing to x-ray mass 
discrepancy in Abell 2218 most likely requires   
a somewhat higher thermal pressure due to the  
hotter temperature of the gas in the core region, the presence of 
significant nonthermal pressure support due to turbulent velocities and 
perhaps amplified magnetic fields, and possibly the chance projection of 
galaxy-group-sized substructure along the line of sight. These effects are 
expected to be 
important at an intermediate stage of a major merger 
($\gtrsim 1$~Gyr after the primary collision) when the return to 
hydrostatic equilibrium in the central region of the cluster has been 
delayed by additional interactions with the subcluster dark matter core.
Better quantitative tests of this picture will be provided by deeper x-ray
imaging, by plasma velocity maps obtained from high-resolution x-ray
spectroscopy, and by better simulations that take into 
account a range of subcluster to primary cluster mass ratios and collision
parameters with sufficient resolution to model physical processes on the 
kiloparsec scales now shown by Chandra observations to be dynamically 
important. 


\section{Summary}
\label{sec:conclude}

In this paper we presented the results from two short observations 
(combined exposure of $\sim 17$~ks) of galaxy cluster Abell 2218 taken 
on October 19, 1999 using the ACIS-I detector on board the Chandra X-ray 
Observatory.  In summary:
\begin{enumerate}

\item{Using a Raymond-Smith single temperature plasma model corrected for 
galactic absorption on data within a $5'.1$ ($\sim 1$~Mpc) radius circular 
aperture of the cluster core, we find a mean cluster temperature of 
$kT = 6.9 \pm 0.5$~keV, metallicity of $0.20 \pm 0.13$, and rest 
frame $2$--$10$~keV luminosity of $6.2 \times 10^{44}$~\ergs 
in the $\Lambda$CDM, $H_0=65$~\kmsmpc cosmology in agreement with previous 
measurements.}

\item{The broadband ($0.3$--$7$~keV) x-ray brightness distribution within 
$4'.2$ of the cluster core is well fit by a simple spherical beta model with
core radius $66''.4$($138h^{-1}$~kpc), index $\beta = 0.70$, and 
normalization $S_{x0} = 0.967$~counts/arcsec$^2$. 
However, the centroid of the distribution
is displaced by $\sim 22''$ to the east and south of the dominant cD galaxy
along the direction of the primary mass concentration found by AbdelSalam, 
Saha \& Williams (1998) and similar to displacements found in analyses of 
ROSAT data by Markevitch (1997) and Allen (1998). If we relax the assumption
of spherical symmetry we find the cluster ellipticity is $0.179$, 
close to spherical, with no significant change in $\beta$ or the location of 
the emission centroid.}

\item{Examination of the central $2'$ of the cluster core reveals that the
peaks of the soft ($0.3$--$2$~keV) and hard ($2$--$7$~keV) x-ray emission 
are displaced from each other (and the emission centroid) along the direction
of the mass elongation of the cluster. Such multi-peaked behavior 
suggests on-going merging activity within the core.}

\item{The gas temperature profile, projected onto a line extending along 
the direction of the mass elongation, shows a temperature increase in the 
region surrounding the dominant cD galaxy and asymmetric behavior with respect
to the emission centroid. Temperatures are close to the cluster mean 
temperature out to $2'$ to the south and east of the emission centroid, while
the temperature falls sharply $1'$--$2'$ to the north and west. }

\item{We see no evidence for strong shocks within the cluster core. However, 
weak shocks can not be ruled out by our data and there may be weak evidence
for cold clouds or trapped gas denoting a complicated core structure.}

\item{The discrepancy of a factor of $2$--$3$ in the lensing to x-ray mass
ratio persists for the smallest critical radii $b \sim 40h^{-1}$~kpc
($21''$--$22''$) measured relative to the dominant cD galaxy.  
Although a chance alignment of substructure along the line of sight could 
reduce the size of this discrepancy, reconciliation of the lensing
and x-ray masses at this critical radius would still require the ratio of 
nonthermal to thermal pressure to be $\gtrsim 1$.  Agreement improves, 
however, for larger $b$ and becomes good for comparisons 
between the weak lensing masses and the enclosed projected x-ray mass 
for critical radii $\gtrsim 200h^{-1}$~kpc($100''$).}

\end{enumerate}

These data, although photon count limited, clearly suggest that the 
intra-cluster gas in the core of Abell 2218 is not in hydrostatic 
equilibrium. The superb spatial resolution of 
the Chandra satellite now makes it feasible to study the details of the 
merging processes expected to be responsible for the formation 
and evolution of galaxy clusters. Our data are qualitatively consistent 
with the view that we are seeing the cluster 
intermediate between 
the primary collision and 
its eventual relaxation to hydrostatic 
equilibrium. Simulations suggest that at this stage of merger evolution gas 
mixing is incomplete, bulk and turbulent flows are present and localized 
magnetic field amplification within the core is maximized.   
With the longer 
exposure observations planned for Abell 2218 in the near term we should be 
able to make a high resolution temperature map of the central regions of the 
cluster to test this hypothesis further. In the future the high resolution 
spectroscopy planned for ASTRO-E2 and Constellation X will allow us to 
map gas velocity flows as well with the hope of finally disentangling 
the dynamics of these complex processes evolving deep within the cluster 
core.  

\acknowledgements

This work is supported in part by NASA contracts NAS-8-37716, NAS-8-38252,
and 1797-MIT-NA-A-38252.  We gratefully thank Leon Van Speybroeck for 
sharing his data, Lilya Williams for making her mass map available to us,
Greg Bryan and Paul Schechter for useful discussions, and 
John Arabadjis for help with the images.  



\clearpage
\plotone{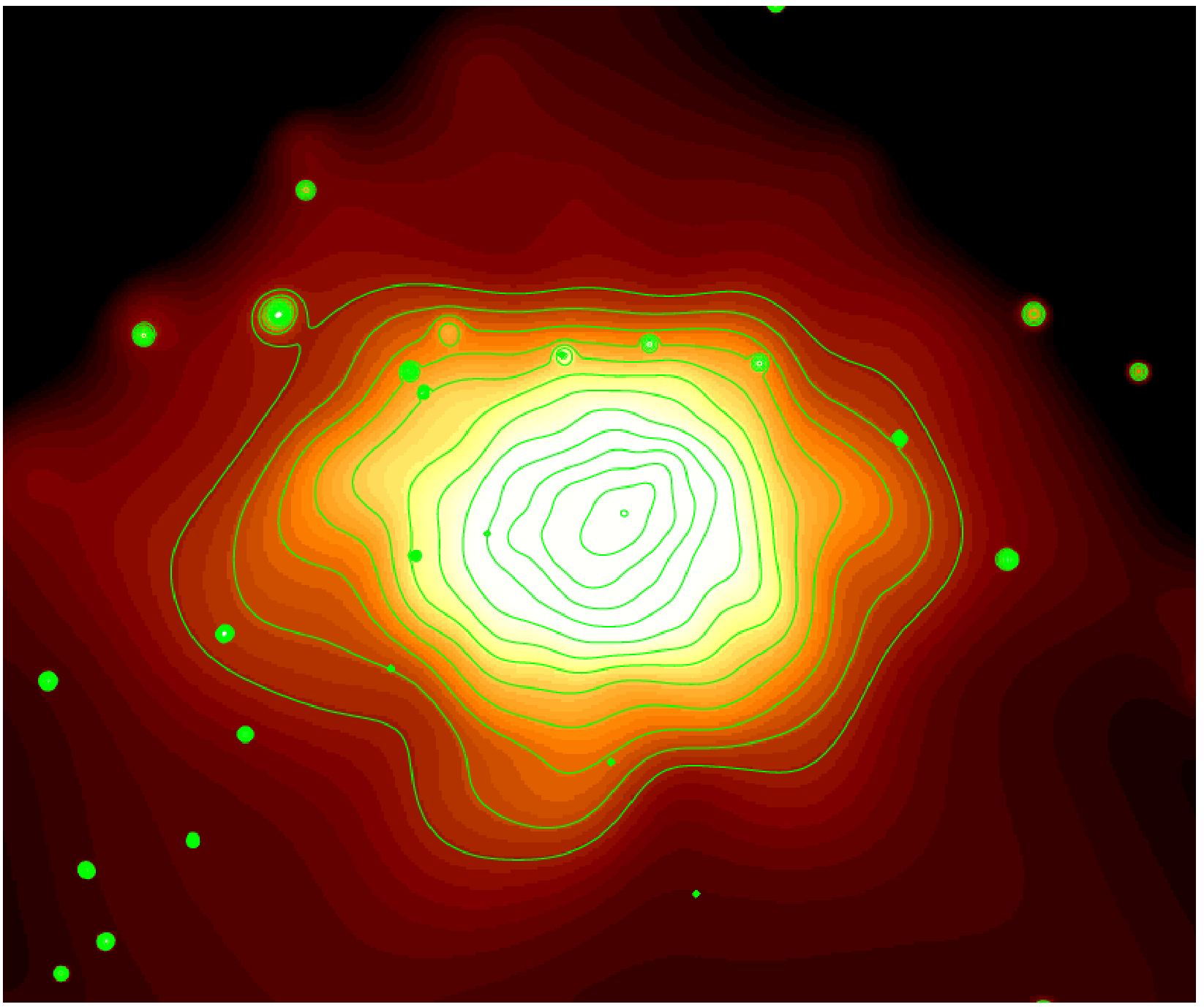}

\figcaption{
Adaptively smoothed, exposure corrected, broadband ($0.3$ - $7$~keV) 
Chandra image of A2218. The field is $14'.2 \times 11'.87$ with north up
and east to the left. Contours range in steps of $\protect\sqrt 2$ from 
$0.0156$~counts/arcsec$^2$ where the lowest contour is twice the background.
\label{fig_fullwpoints}}

\clearpage
\plotone{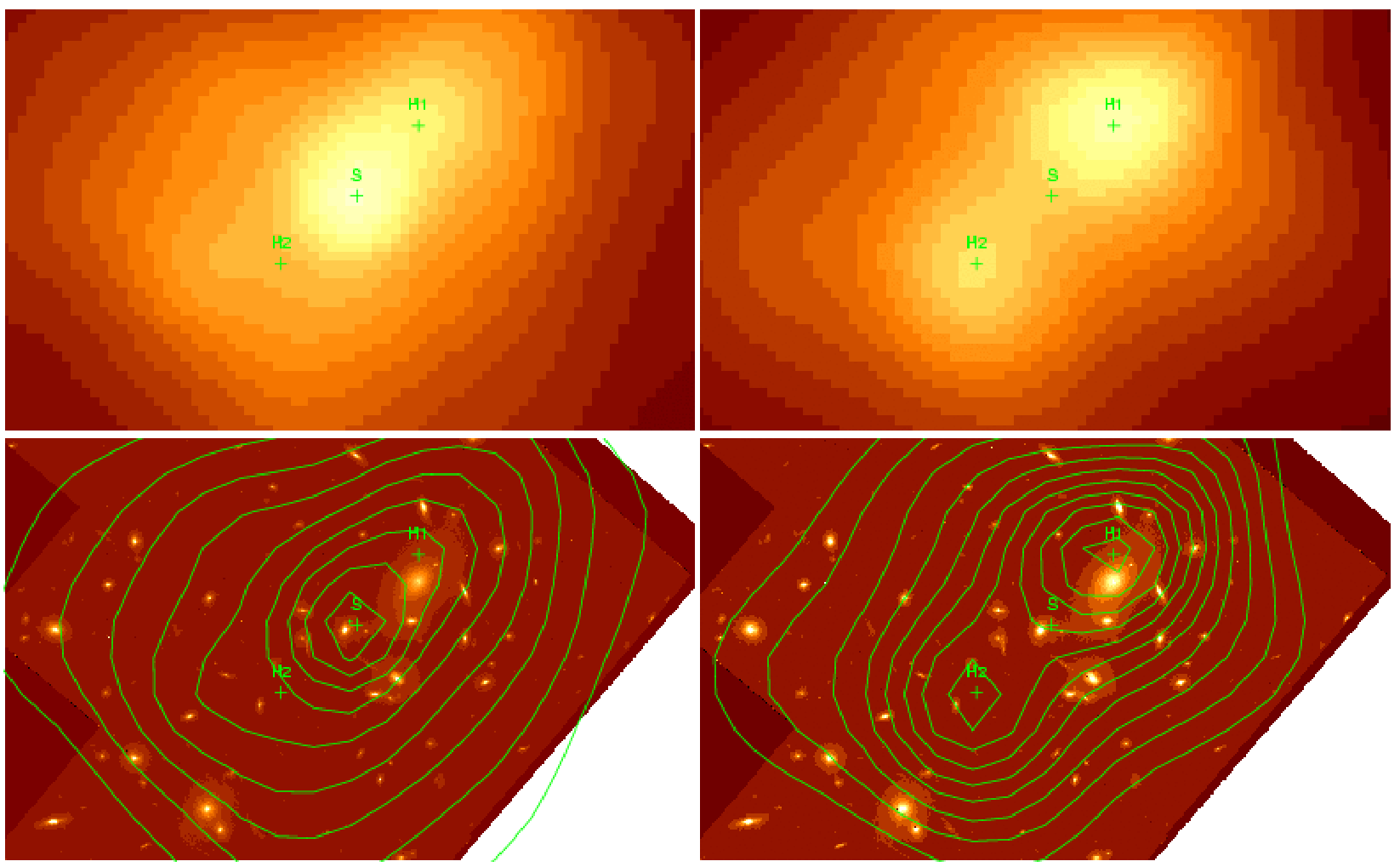}

\figcaption{Adaptively smoothed, exposure corrected images of soft 
($0.3$ - $2$~keV; upper left) and hard ($2$ - $7$~keV; upper right) emission 
from the core of Abell 2218. The  soft (hard) peaks are denoted by 
S (H1,H2), respectively. In the lower left(right) images,  
contours from the soft(hard) band x-ray images are superposed on the Hubble 
image of the same field.
\label{fig_corecsmoothimg}}

\clearpage
\plotone{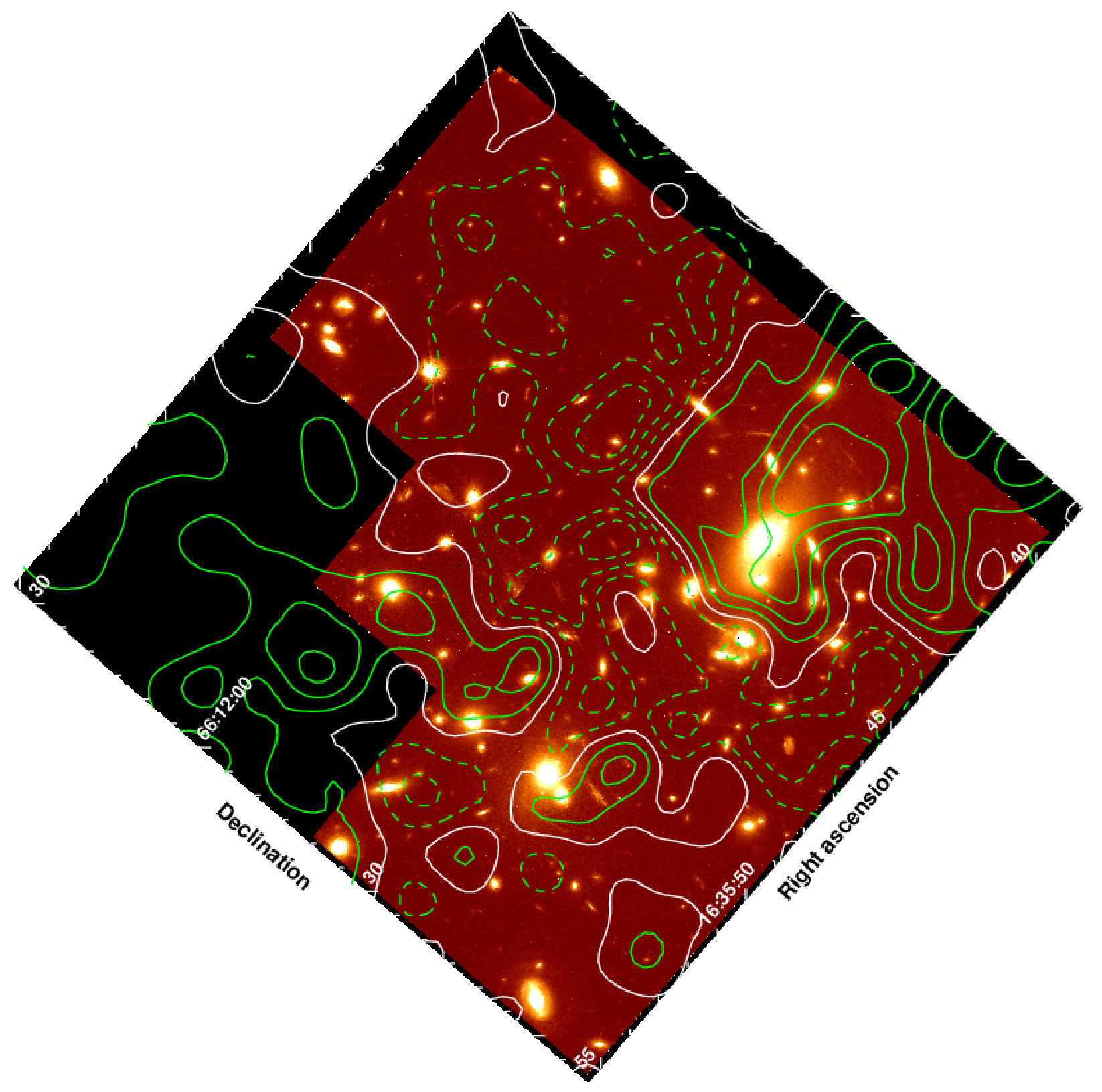}

\figcaption{
Positive (solid green), zero (solid white) and negative (dashed green) 
contours from the gaussian smoothed ($\sigma=5''$) image of the 
residuals from the large field beta model fit to the $0.3$ - $7$~keV x-ray
surface brightness overlaid on the Hubble mosaic optical image of the core 
of Abell 2218.
\label{fig_bbmodelresiduals}}

\clearpage
\plotone{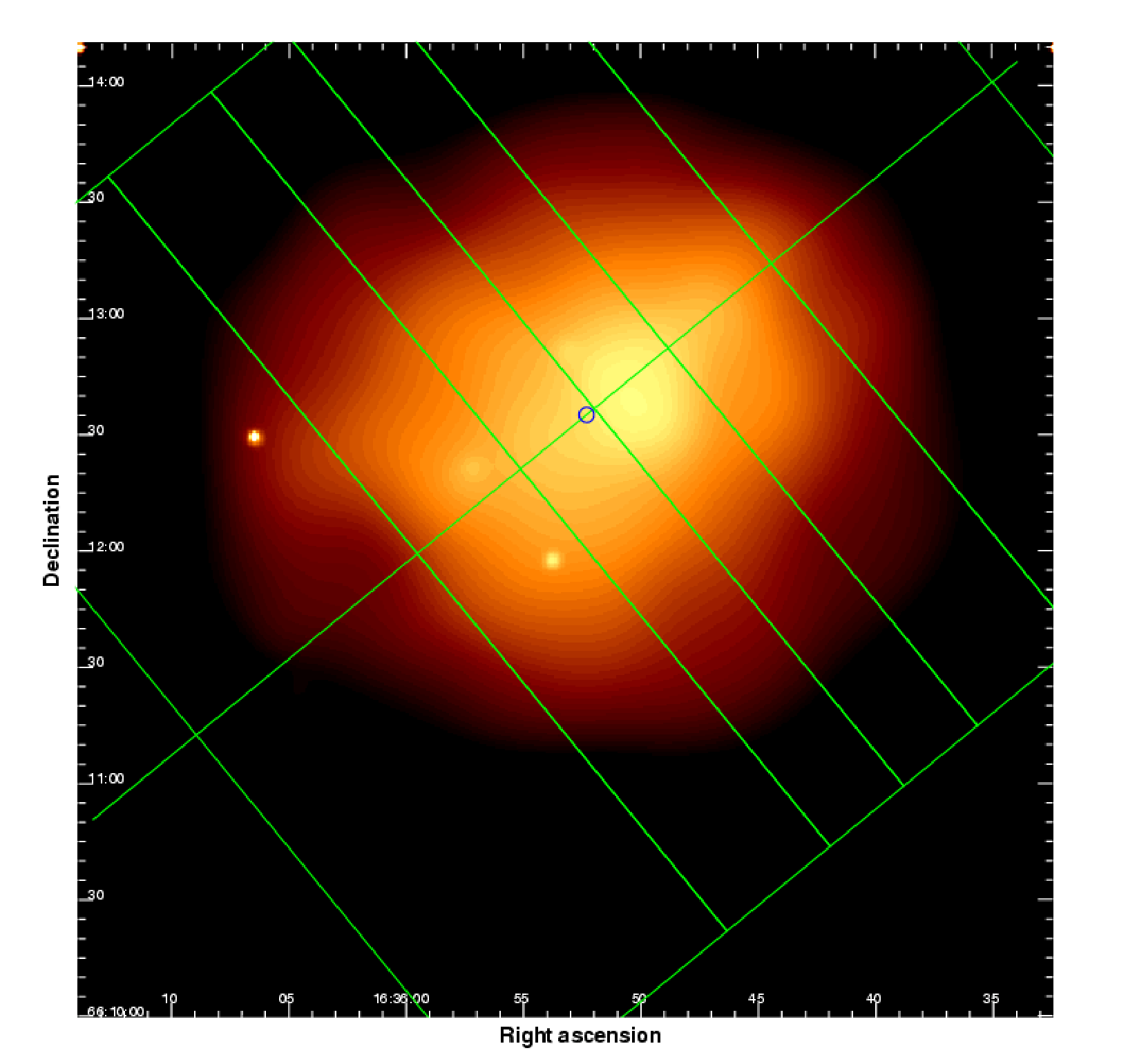}

\figcaption{
Broadband ($0.3$ - $7$~keV) adaptively smoothed image of the inner 
$4'.2 \times 4'.2$ core of the cluster overlaid with the axis line and
temperature grid used in Figure \protect\ref{fig_btprofiles}. The mean beta
model centroid ($r=0$ in Figure \protect\ref{fig_btprofiles}) is denoted by
a circle.
\label{fig_bbaxisgrid}}

\clearpage
\plotone{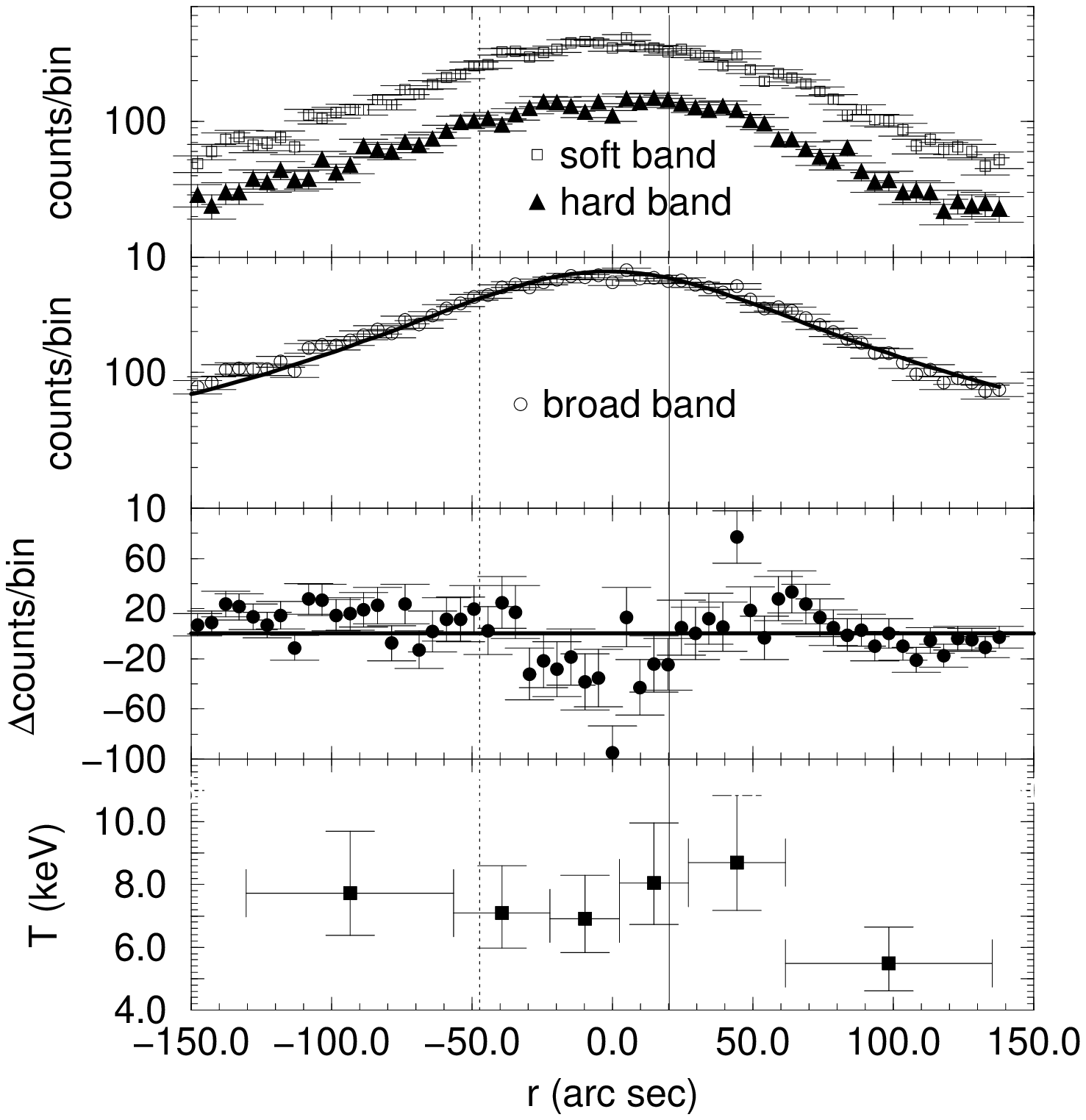}

\figcaption{
(a)Surface brightness profiles for the soft band ($0.3$ - $2$~keV, open 
squares) and hard band ($2$ - $7$~keV, filled triangles) projected on the 
axis 
shown in Figure \protect\ref{fig_bbaxisgrid} with the 
origin placed at the mean x-ray distribution centroid. Error bars are 
$\protect\sqrt N$ Poissonian errors only and the bin width along the line is 
$4''.92$. Vertical solid (dotted) lines denote the projected positions of 
the dominant cD (second brightest elliptical) galaxies, respectively. 
(b)Surface brightness profile for the broadband ($0.3$ - $7$~keV) 
using the same coordinate conventions as in (a). The solid line is 
the best fit beta model for the broadband data. (c) Residuals between the 
broadband surface brightness distribution and the large field beta model 
projected onto the symmetry line. Errors are $\protect\sqrt N$ only. 
(d)Temperature profile through the core. The error bars in position denote 
the bin width along the symmetry line.  Temperature errors are 
$90\protect\%$~CL from the fit.
\label{fig_btprofiles}}

\clearpage
\plotone{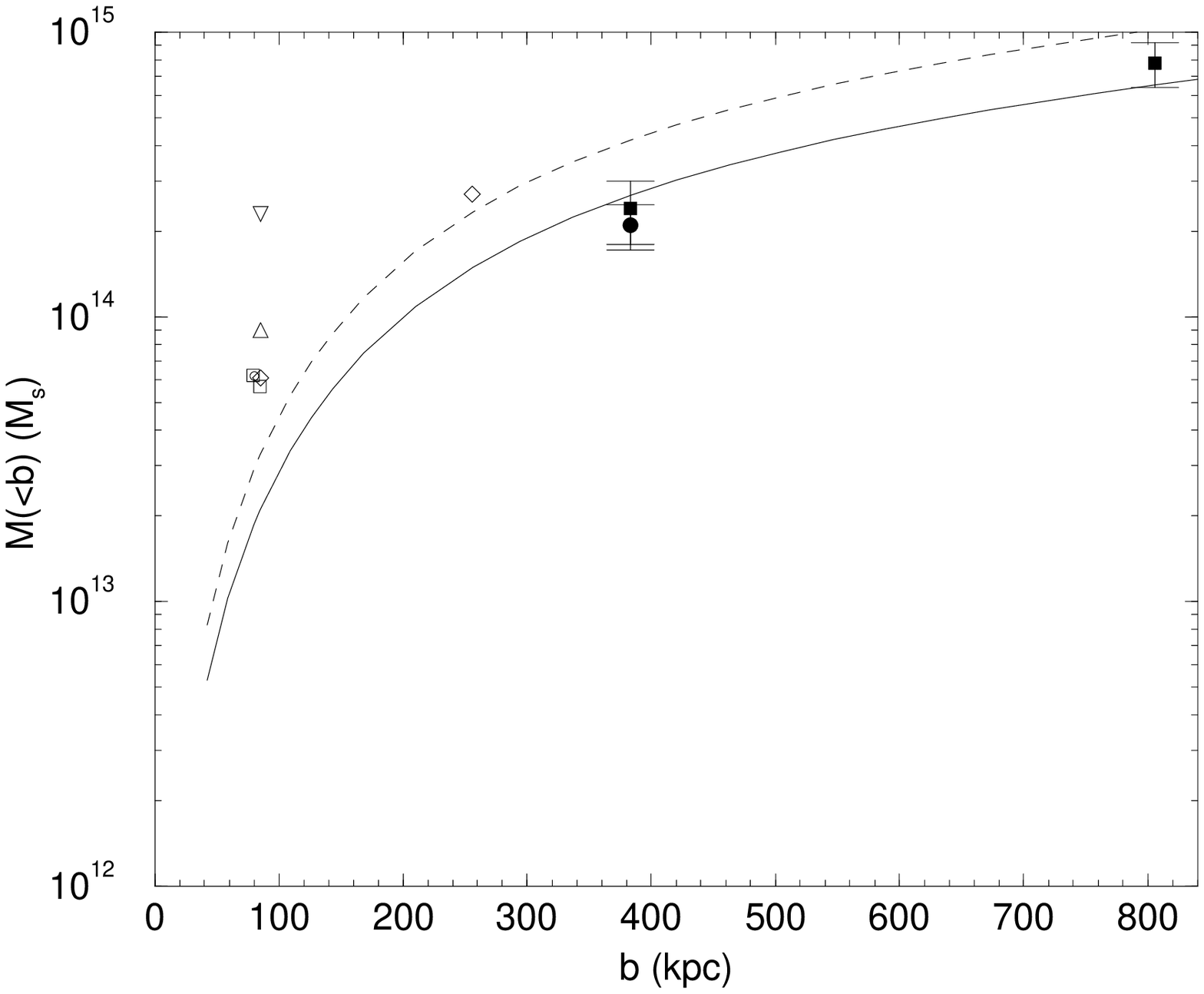}

\figcaption{
Projected total x-ray mass as a function of radial distance $b$
from the dominant cD galaxy using an isothermal spherical beta model with 
temperatures $kT=6.9$~keV (solid line) and $kT=10.8$~keV (dashed line), 
respectively, and the sCDM cosmology ($\Omega_m=1$,$H_0=50$~\kmsmpc). The 
strong lensing results are from Loeb \& Mao 1994 (open circle), 
Kneib \etal 1995 (open diamonds) and Allen 1998 (open boxes). Weak lensing 
results are from Squires, \etal 1996 (filled squares) and Smail \etal 1997 
(filled circle).  Triangle up (down) are the minimum (maximum) mass 
reconstructions of AbdelSalam \etal 1998 using both weak and strong lensing 
constraints.
\label{fig_massprofile}}
 

\clearpage

\begin{table}
\centerline{\begin{tabular}{|c|c|c|c|c|} \hline
Energy band & $kT$ & Abundance & $L_x$($2-10$~keV)  
& $\chi^2/dof$ \\ 
 (keV) & (keV) &  & (\ergs) & \\ \hline \hline
$0.7$--$9$ & $6.9 \pm 0.5$  & $0.20 \pm 0.13$ & 
$6.2 \times 10^{44}$ & $433/436$ \\ \hline
\end{tabular}}
\caption{ Mean spectral properties of the cluster using an aperture of 
radius $5'.1$ ($0.64h^{-1}$~Mpc) on CCD I-3 assuming Galactic absorption, 
$3 \times 10^{20}$~\cms. Errors are $90\%$ CL.}
\label{tab:meanprop}
\end{table}

\begin{table}
\centerline{\begin{tabular}{|c|c|c|c|c|c|c|} \hline
Field & Energy band & Centroid  & $r_c$ & $\beta$ &
$S_{x0}$  & $\epsilon$ \\
 & (keV) & $\alpha$($^h$ $^m$ $^s$), $\delta$($^\circ$ $'$ $''$) & $('')$ &  & 
(counts/arcsec$^2$) & \\  
\hline
\hline
mean & $0.3$--$7$ & 
$16\,\,35\,\,52.4$, $66\,\,12\,\,34.3$ & 
$66.4 \pm 0.45$ & $0.705^{+0.004}_{-0.003}$ & $0.965 \pm 0.012$ & $0$ \\
\hline
mean & $0.3$--$7$ & 
$16\,\,35\,\,52.4$, $66\,\,12\,\,34.3$ & 
$73.1_{-0.4}^{+0.6}$ & $0.706^{+0.003}_{-0.004}$ &
$0.976 \pm 0.012$ & $0.18 \pm 0.01$ \\ \hline 
core & $0.3$--$7$ & 
$16\,\,35\,\,52.3$, $66\,\,12\,\,35.4$ & 
$75.1 \pm 0.6$ & $0.774^{+0.006}_{-0.007}$ & $0.928 \pm 0.012$ & $0$ \\ \hline
core & $0.3$--$2$ & 
$16\,\,35\,\,52.2$, $66\,\,12\,\,35.2$ & 
$77.6 \pm 0.8$ & $0.792^{+0.008}_{-0.007}$ & $ 0.668 \pm 0.011$ & $0$ \\ 
\hline
core & $2$--$7$ & 
$16\,\,35\,\,52.3$, $66\,\,12\,\,36.1$ & 
$69.8 \pm 1.2$ & $0.737^{+0.012}_{-0.012}$ & $0.260 \pm 0.007$ & $0$ \\ \hline
\end{tabular}}
\caption{ Beta model fits of the cluster where the centroid of the beta model 
given in J2000 coordinates has been determined by the fit. The mean (core)
fields are $8'.4 \times 8'.4$ ($4'.2 \times 4'.2$) centered on the peak of
the broadband emission. Other fit parameters are defined as follows:  
$r_c$ is the core radius; 
$\beta$ is the index given in Equation \protect\ref{eq:sbeta}; 
$S_{x0}$ is the central brightness; $\epsilon$ is the ellipticity. 
Errors are $90\%$ confidence limits. Formal errors for the centroid are 
$\lesssim 2''$. }
\label{tab:betacore}
\end{table}
 
\begin{table}
\centerline{\begin{tabular}{|c|c|c|c|} \hline
Feature & Energy band & Offset & Offset  \\ 
 & (keV) & $('')$ & ($h^{-1}$kpc) \\ 
\hline \hline
centroid & $0.3-7$ & $-21.9$  & $-45.6$ \\ \hline
peak     & $0.3-7$ & $-9.5$   & $-19.8$ \\ \hline
peak(S)  & $0.3-2$ & $-17.5$  & $-36.4$ \\ \hline
peak(H1) & $2-7$   &  $+4.5$  & $+9.4$   \\ \hline
peak(H2) & $2-7$   & $-39.2$  & $-81.6$ \\ \hline
\end{tabular}}
\caption{ Offsets of the x-ray features from the dominant cD galaxy. Negative
(positive) offsets are to the south and east (north and west), respectively.
Peak positions are taken from adaptively smoothed images.
 }
\label{tab:offset}
\end{table} 

\begin{table}
\centerline{\begin{tabular}{|c|c|c|c|c|c|c|c|} \hline
Lensing Reference & $b$ & $b$ & $M_{lens}$ & 
$M_{xc}$ & $M_{xh}$ & $M_{lens}/M_{xc}$ &  $M_{lens}/M_{xh}$ \\ 
   & $('')$ & (kpc) & $(10^{14} M_{\odot})$ & $(10^{14} M_{\odot})$ &
 $(10^{14} M_{\odot})$ &  &  \\
\hline
\hline
 Loeb \& Mao 1994 & $20.8$ & $79.8$ & $0.64$ & $0.19$ & $0.29$  & $3.4$ & 
$2.2$ \\ \hline
 Allen 1998 & $20.7$ & $79.4$ & $0.62$ & $0.19$ & $0.29$ & $3.4$ & $2.1$ \\ 
  \hline
 Kneib \etal 1995 & $22.1$ & $84.8$ & $0.61$ & $0.21$  & $0.32$  & $2.9$ & 
  $1.9$ \\ \hline
 Allen 1998 & $22.1$ & $84.8$ & $0.57$ & $0.21$  & $0.32$ & $2.7$ & $1.8$ \\ 
\hline
 AbdelSalam \etal 1998 & $22.1$ & $84.8$ & $0.9$ & $0.21$ & $0.32$ & $4.3$ &
  $2.8$ \\ \hline
 Kneib \etal 1995 & $66.7$ & $256$ & $2.7$ &$1.49$ & $2.33$ & $1.8$ & $1.2$ \\
 \hline
 Squires \etal 1996 & $99.8$ & $383$ & $2.4 \pm 0.6$ & $2.67$ & $4.18$ & 
 $0.9$ & $0.6$ \\ \hline
 Smail \etal 1997 & $99.8$ & $383$ & $2.1 \pm 0.38$ & $2.67$ & $4.18$ & $0.8$ 
& $0.5$ \\ \hline
 Squires \etal 1996 & $210$ & $806$ & $7.8 \pm 1.4$ & $6.52$ & $10.20$ & 
 $1.2$  & $0.8$ \\ \hline      
\end{tabular}}
\caption{ Lensing to X-ray Mass Ratios. The first five entries are strong 
lensing results, while the last three are lower bounds on the mass 
from weak lensing analyses. 
$M_{xc}$ 
and $M_{xh}$ are the x-ray masses for $kT=6.9$~keV and the upper bound
$kT=10.8$~keV, respectively, assuming an $\Omega_m=1$, $\Omega_\Lambda=0$, 
$h=0.5$ cosmology. The minimum mass model result is used from AbdelSalam 
\etal (1998).}
\label{tab:massratio}
\end{table}


\begin{thebibliography}{it}

\bibitem[AbdelSalam \etal (1998)]{abdel}
        AbdelSalam, H.M., Saha, P., \& Williams, L.L.R. 1998, \aj, 116, 1541
 
\bibitem[Allen (1998)]{allen98}
        Allen, S.W. 1998, \mnras, 296, 392

\bibitem[Bartlemann \& Steinmetz (1996)]{bart96}
        Bartelmann, M. \& Steinmetz, M. 1996, \mnras, 283, 431

\bibitem[Bautz \etal (1998)]{acismwb98} 
        Bautz, M.W., \etal 1998, X-ray Optics, Instruments and Missions, ed.
        R.B. Hoover \& A. B. Walker, Proc. SPIE, 3444, 210
       
\bibitem[Birkinshaw \& Hughes (1994)]{birk94}
        Birkinshaw, M. \& Hughes, J.P. 1994, \apj, 420, 33

\bibitem[Bryan (1996)]{gbryan96}
        Bryan, G.L 1996, Ph.D. Thesis, University of Illinois at 
        Urbana-Champaign

\bibitem[Buote (2001)]{buote01}
        Buote, D. 2001, \apjl, in press, preprint astro-ph/0104211

\bibitem[Buote \& Tsai (1996)]{buotetsai}
        Buote, D.A. \& Tsai, J.C. 1996, \apj, 458, 27
 
\bibitem[Clarke \etal (2001)]{clarke}
        Clarke, T.E., Kronberg, P.P. \& Bohringer, H. 2001, \apjl, submitted,
        preprint astro-ph/0011281

\bibitem[Cannon \etal (1998)]{cannon}
        Cannon, D.B., Ponman, T.J \& Hobbs, I.S. 1998, \mnras, 302, 9

\bibitem[David \etal (1993)]{david93}
        David, L.P., Slyz, A., Jones, C., Forman, W.,Vrtilek, S.D., \&
        Arnaud, K.A. 1993, \apj, 412, 479

\bibitem[Dobrzycki \etal (1999)]{dob99}
        Dobrzycki, A. Ebeling, H., Glotfelty, K., Freeman, P., Damiani, F.,
        Elvis, M., \& Calderwook, T. 1999, Chandra Detect v$1.0$ User's 
        Guide, Chandra X-ray Center

\bibitem[Dupke \& Bregman (2001)]{dupke}
        Dupke, R.A. \& Bregman, J.N. 2001, preprint astro-ph/0106030

\bibitem[Evrard (1990)]{evrard90}
        Evrard, A.E. 1990, \apj, 363, 349
        
\bibitem[Evrard \etal (1996)]{evrard}
        Evrard, A.E., Metzler, C.A. \& Navarro, J.F. 1996, \apj, 469, 494

\bibitem[Freeman \etal (2001)]{free01}
        Freeman, P.E., Kashyap, V., Rosner, R., \& Lamb, D.Q. 2001, \apj, 
        submitted
 
\bibitem[Garmire \etal (1992)]{gamire92}
        Garmire, G.P. \etal 1992, AIAA, Space Programs and Technologies 
        Conference, Huntsville, AL, March 24-27

\bibitem[Giovanni \etal (1999)]{giovanni99}
        Giovanni, G., Tordi, M. \& Feretti, L. 1999, New Astronomy, 4, 141

\bibitem[Gomez \etal (2000)]{gomez}
        Gomez, P.L., Loken, C., Roettiger, K. \& Burns, J.O. 2000, preprint
        astro-ph/0009465

\bibitem[Kneib \etal (1995)]{kneib95}
        Kneib, J.P., Mellier, Y., Pello, R., Miralda-Escude, J., 
        Le Borgne, J.-F., Bohringer, H., Ficat, J.-P. 1995, A \& A, 303, 27

\bibitem[Kneib \etal (1996)]{kneib96} 
        Kneib, J.P., Ellis, R.S., Smail, I., Couch, W., \& Sharples, R.M.
        1996, \apj, 471, 643

\bibitem[Kraft \etal (1991)]{kraft91}
        Kraft, R.P., Burrows, D.N. \& Nousek, J.A 1991, \apj, 374, 344
    
\bibitem[Le Borgne \etal (1992)]{leborg92}
        Le Borgne, J.F., Pello, R. \& Sanahuja, B. 1992, A \& AS, 95, 87

\bibitem[Loeb \& Mao (1994)]{loeb94}
        Loeb, A. \& Mao, S. 1994, \apj, 435, L109

\bibitem[Loewenstein (1997)]{loew97}
        Loewenstein, M. 1997, in X-Ray Imaging and Spectroscopy of Cosmic 
        Plasmas, ed. F. Makino (Tokyo: Universal Academy), 67

\bibitem[Makino (1996)]{makino96}
        Makino, N. 1996, PASJ, 48, 573

\bibitem[Markevitch (1997)]{mvitch97}
        Markevitch, M. 1997, \apj, 483, L17

\bibitem[Markevitch \etal (2000)]{mvitch00}
        Markevitch, M., Vikhlin,  A, Mazzotta, P., VanSpeybroeck, L. 2000,
        preprint astro-ph/ 0012215

\bibitem[Markevitch (2001)]{mvitch01}
       Markevitch, M. 2001, \hfil \newline  
 http://asc.harvard.edu/cal/Links/Acis/acis/Cal\_prods/bkgrnd/current/index.html

\bibitem[Mazzotta \etal (2001)]{mazz01}
        Mazzotta, P., Markevitch, M., Vikhlinin, W.R., Forman, L., David, P.,
        \& VanSpeybroeck, L. 2001, preprint astro-ph 0102291

\bibitem[McHardy \etal (1990)]{mchardy90}
        McHardy, I.M., Stewart, G.C., Edge, A.C., Cooke, B., Yamashita, K. 
        \& Hatsukade, I. 1990, \mnras, 242, 215

\bibitem[Mihalas \& Mihalas (1984)]{mihalas}
        Mihalas, D. \& Mihalas, B.W. 1984, Foundations of Radiation 
        Hydrodynamics (Oxford University Press, Oxford), 230--239

\bibitem[Miralda-Escude \& Babul (1995)]{jordi}
        Miralda-Escude, J. \& Babul, A. 1995, \apj, 449, 18

\bibitem[Moffet \& Birkinshaw (1989)]{moffet}
        Moffet, A.T. \& Birkinshaw, M. 1989, \aj, 98, 1148

\bibitem[Monet (1998)]{monet98}
        Monet, D.G. 1998, BAAS, 193, 120.03

\bibitem[Morrison \& McCammon (1983)]{wabs}
        Morrison \& McCammon 1983, \apj, 270, 119

\bibitem[Mushotzky \& Loewenstein (1997)]{mushloew97}
        Mushotzky, R.F. \& Loewenstein, M. 1997, \apj, 481, L63

\bibitem[Mulchaey \& Zabludoff (1998)]{mulch98}
        Mulchaey, J.S \& Zabludoff, A.I. 1998, \apj, 496, 73
      
\bibitem[Natarajan \& Kneib (1997)]{nata96}
        Natarajan, P. \& Kneib, J.-P. 1997, \mnras, 283, 1031

\bibitem[Pello \etal (1992)]{pello92}
        Pello, R., Le Borgne, J.F., Sanahuja, B., Mathez, G., \& Fort, B.
        1992, A \& A, 266, 6

\bibitem[Prigozhin \etal (2000)]{prig00}
        Prigozhin, G.,Kissel, S., Bautz, M., Grant, C., LaMarr, B., Forster, 
        R., Richer, G., \& Gamire, G. 2000, Proc. SPIE, 4012, 720

\bibitem[Raymond \& Smith (1977)]{raymond}
        Raymond \& Smith 1977, \apjs, 35, 419

\bibitem[Roettiger \etal (1998)]{roettiger98}
        Roettiger, K., Stone, J.M. \& Mushotzky, R.F. 1998, \apj, 493, 62

\bibitem[Roettiger \etal (1999)]{roettiger99}
        Roettiger, K., Stone, J.M. \& Burns, J.O. 1999, \apj, 518, 594.

\bibitem[Sarazin (1986)]{sarazin86}
        Sarazin, C.L. 1986, Rev. Mod. Phys., 58, 1

\bibitem[Sarazon (2000)]{sarazin00}
        Sarazin, C.L. 2000, in Constructing the Universe with 
        Clusters of Galaxies, ed. F. Durret \& D. Gerbel, in press,
        preprint astro-ph/0009094

\bibitem[Sarazin (2001)]{sarazin01}
        Sarazin, C.L. 2001, The Physics of Cluster Mergers in Merging 
        Processes in Clusters of Galaxies, ed. L. Feretti, Gioia, and
        G. Giovannini (Dordrecht: Kluwer), in press, preprint 
        astro-ph/0105418

\bibitem[Siddiqui (1995)]{sidd95}
        Siddiqui, H. 1995, Ph.D. thesis, University of Leicester

\bibitem[Smail \etal (1997)]{smail97}
        Smail, L, Ellis, R.E., Dressler, A., Couch, W.J., Oemler, A., 
        Sharples, R.M., Butcher, H. 1997, \apj, 479, 70
        
\bibitem[soucail \etal (2000)]{ota00}
        Soucail, G., Ota, N., Bohringer, H., Czoske, O., Hattori, M., \&
        Mellier, Y. 2000, A \& A, 355, 433
 
\bibitem[Squires \etal (1996)]{squires96}
        Squires, G., Kaiser, N., Babul, A., Fahlman, G., Woods, D, 
        Neumann, D.M., \& Boehringer, H. 1996, \apj, 461, 572

\bibitem[Sunyaev \& Zel'dovich (1972)]{sunzeld72}
        Sunyaev, R.A. \& Zel'dovich, Y.B. 1972, Comm. Astrophys. Space 
        Phys., 4, 173

\bibitem[Sunyaev \& Zel'dovich (1980)]{sunzeld80}
        Sunyaev, R.A. \& Zel'dovich, Y.B. 1980, ARA \& A, 18, 537

\bibitem[Vikhlinin \etal (2001a)]{vik01a}
        Vikhlinin, A., Markevitch, M., \& Murray, S.M. 2001, \apj, 549, L47

\bibitem[vikhlinin \etal (2001b)]{vik01b}
        Vikhlinin, A., Markevitch, M., \& Murray, S. M. 2001, \apj, 551, 160 

\bibitem[Zabludoff \& Mulchaey (1998)]{zab98}
        Zabludoff, A.I. \& Mulchaey, J.S. 1998, \apj, 496, 39 

\bibitem[Ziegler \etal (2001)]{zeigler01}
        Ziegler, B.L., Bower, R.G., Smail, I., Davies, R.L. \& Lee, D. 
        2001, \mnras, in press, preprint astro-ph/0103475
        
\end{thebibliography}
\end{document}